\documentclass[12pt]{iopart}
 
\usepackage{colortbl}
\usepackage{graphicx}% Include figure files
\usepackage{dcolumn}% Align table columns on decimal point
\usepackage{bm}% bold math
\usepackage{hyperref}% add hypertext capabilities
\usepackage{multirow}
\usepackage{multicol}
\usepackage{supertabular}
\usepackage{inputenc}
\usepackage{cite}

\hypersetup{pdfborder=0 0 0,colorlinks=true,citecolor=blue,linkcolor=blue}

\newcommand{\cspbiRpSingle}{Cs$_2$PbI$_4$}
\newcommand{\cspbiRpDouble}{Cs$_3$Pb$_2$I$_7$}
\newcommand{\cspbbrRpSingle}{Cs$_2$PbBr$_4$}
\newcommand{\cspbbrRpDouble}{Cs$_3$Pb$_2$Br$_7$}
\newcommand{\cspbclRpSingle}{Cs$_2$PbCl$_4$}
\newcommand{\cspbclRpDouble}{Cs$_3$Pb$_2$Cl$_7$}
\newcommand{\minus}[1]{$\overline{#1}$}
\newcommand{\polColorRow}{\rowcolor[rgb]{ .776,  .878,  .706}}
\newcommand{\mycc}{\cellcolor[rgb]{.776, .878, .706}}

\newcommand{\glazer}[6]{#1$^#2$#3$^#4$#5$^#6$}

\begin{document}

\title[]{Emergence of Rashba-/Dresselhaus Effects in Ruddlesden-Popper Halide Perovskites with Octahedral Rotations}

\author{Sonja Krach}
\address{Institute of Physics, University of Bayreuth, 95440 Bayreuth, Germany}
\author{Nicol\'as Forero-Correa}
\address{Doctorado en Fisicoqu\'imica Molecular, Facultad de Ciencias Exactas, Universidad Andres Bello, Santiago 837-0136, Chile} 
\author{Raisa-Ioana Biega}%
\address{MESA+ Institute for Nanotechnology, University of Twente, 7500 AE Enschede, The Netherlands}
\author{Sebastian E. Reyes-Lillo}
\address{Departamento de Ciencias F\'isicas, Universidad Andres Bello, Santiago 837-0136, Chile}
\author{Linn Leppert}
\address{MESA+ Institute for Nanotechnology, University of Twente, 7500 AE Enschede, The Netherlands}
\ead{l.leppert@utwente.nl}

\date{\today}

\begin{abstract}
	Ruddelsden-Popper halide perovskites are highly versatile quasi-two-dimensional energy materials with a wide range of tunable optoelectronic properties. Here we use the all-inorganic Cs$_{n+1}$Pb$_n$X$_{3n+1}$ Ruddelsden-Popper perovskites with X=I, Br, and Cl to systematically model the effect of octahedral tilting distortions on the energy landscape, band gaps, macroscopic polarization, and the emergence of Rashba-/Dresselhaus splitting in these materials. We construct all unique $n=1$ and $n=2$ structures following from octahedral tilts and use first-principles density functional theory to calculate total energies, polarizations and band structures, backed up by band gap calculations using the $GW$ approach. Our results provide design rules for tailoring structural distortions and band-structure properties in all-inorganic Ruddelsden-Popper perovskites through the interplay of the amplitude, direction, and chemical character of the antiferrodistortive distortion modes contributing to each octahedral tilt pattern. Our work emphasizes that, in contrast to 3D perovskites, polar structures may arise from a combination of octahedral tilts, and Rashba-/Dresselhaus splitting in this class of materials is determined by the direction and Pb-I orbital contribution of the polar distortion mode.
\end{abstract}

\maketitle

\section{Introduction}
Ruddlesden-Popper (RP) halide perovskites are heterogeneous, layered materials with exceptional photophysical properties due to their quasi-two-dimensional (2D) structures \cite{Mitzi1995,Mitzi1996a, saparov_organicinorganic_2016, Tsai2016, Stoumpos2016a, Smith2018a, Blancon2018, Even2014, mcnulty_structural_2021}. This family of materials with chemical formula A$_{n-1}$A'$_2$B$_n$X$_{3n+1}$ can be thought of as derived from the 3D ABX$_3$ perovskites, where A and A' are monovalent (organic) cations, B is a divalent metal cation (e.g. Pb), X is a halogen anion, and $n$ refers to the number of perovskite layers. Even more so than their 3D congeners, RP perovskites feature an enormous structural versatility and can be synthesized in bulk form \cite{Mitzi1995, Mitzi1996a, Smith2018a, Gao_MolecularEngineering_2019, Connor2018, Connor2020a, Coffey_ControllingCrystalization_2022}, exfoliated as a monolayer \cite{Yaffe2015a}, and assembled in interfaces or heterostructures with other layered materials \cite{aubrey_directed_2021, Shi_TwodimensionalHalide_2020}. They owe this flexibility and their robust stability under ambient conditions in a large part to the wide range of organic A and A' site cations which have been shown to significantly affect the photophysical and thermal properties of this class of materials due to structural and dielectric effects \cite{Pedesseau2016, cortecchia_structure-controlled_2018, Fridriksson2020a, Dhanabalan2020a, Jana2020, Menahem2021, Li_2DRulebook_2021, Filip2022}. 

A particularly intriguing feature of RP perovskites is that their quasi-2D structure, in which the corner-sharing connectivity of the BX$_6$ octahedra is severed in the stacking direction, can lead to a rich landscape of octahedral tilting distortions \cite{mulder_turning_2013}. In 3D metal-halide perovskites, octahedral tilting is the primary energy-lowering structural distortion. It is well-understood that octahedral tilting in 3D metal-halide perovskites leads to a blueshift of the band gap due to the antibonding nature of the valence band maximum (VBM) and conduction band minimum (CBM) \cite{Zhao2021}. However, octahedral tilting in 3D perovskites cannot lead to non-centrosymmetric structures; these distortions preserve a global centrosymmetric space group. 3D perovskites with polar space groups exhibiting properties like a macroscopic polarization, ferroelectriciy, or Rashba-/Dresselhaus splitting are therefore rare \cite{Benedek2013}. This is different in RP perovskites. In RP oxides, octahedral tilting distortions have been explored in depths and led to the development of the concept of hybrid improper ferroelectricity \cite{Benedek2011} followed by a multitude of predictions of ferroelectric oxides \cite{Birol2011, balachandran_crystal-chemistry_2014}. In RP halide perovskites, the energy landscape of octahedral distortions and their effects on the electronic structure properties has not been systematically studied yet.

Halide perovskites have garnered attention due to their potential for spin-dependent physics that arises as a consequence of strong spin-orbit coupling due to the presence of heavy B site metals like Pb \cite{Even2015a, Kepenekian2017}. In combination with noncentrosymmetric structures, spin-orbit coupling leads to the Rashba-/Dresselhaus effect, a lifting of the degeneracy of the energy bands in reciprocal space \cite{Rashba1960, Dresselhaus1955}. This effect can be used for spin-orbitronics, spin-dependent exciton physics, and circular dichroism \cite{Manchon2015, Yang2016e, bourelle_optical_2022, Jana2020} and is tunable through chemical substitution, electric fields and epitaxial strain \cite{Leppert2016}. It has also been speculated to contribute to the long carrier recombination times observed in 3D halide perovskites \cite{Zheng2015}, although the experimental evidence for a Rashba-/Dresselhaus effect in 3D halide perovskites is debated due to the lack of a global noncentrosymmetric structure of these materials \cite{Etienne2016a, Bernardi2018, Davies2018}.

A recent study by Jana \textit{et al.}, combining first-principles calculations and experimental structural characterization unraveled design rules for Rashba-/Dresselhaus splitting in hybrid organic-inorganic 2D halide perovskites \cite{Jana2021}. These authors showed that some quasi-2D hybrid perovskites do not exhibit Rashba-/Dresselhaus splitting, despite their globally noncentrosymmetric space groups. Instead, Jana \textit{et al.} correlated the magnitude of the splitting in reciprocal space to an asymmetry in the interoctahedral tilting angles in the inorganic monolayer. For the case of all-inorganic RP perovskites Cs$_{n+1}$Pb$_n$I$_{3n+1}$ with $n=1,2,3$, Maurer \textit{et al.} showed that only certain types of polar distortions lead to significant Rashba-/Dresselhaus splitting in the CBM \cite{Maurer2022}. While this systematic study clarified that Rashba-/Dresselhaus splitting is non-negligible in RP perovskites with polar distortions comprising diagonal Pb displacements towards the in-plane edges of the PbI$_6$ octahedron, the effect of the energetically more favorable octahedral tilting distortions on Rashba-/Dresselhaus splitting in RP perovskites was not studied.

Here we use first-principles density functional theory (DFT) to systematically study octahedral tilting distortions in all-inorganic Pb-based RP perovskites. We construct all possible octahedral tilting distortions in Cs$_{n+1}$Pb$_n$X$_{3n+1}$ (X=I, Br, Cl) with $n=1$ and $n=2$ layers and determine their effect on the complex energy landscape of these materials, elucidating energetically favorable distortion modes. We then determine how octahedral tilting modify the band gap of these systems, considering both the size of the gap as well as the emergence of Rashba-/Dresselhaus splitting. Our results, supported by many-body perturbation theory calculations, show which distortion modes have the most significant effects on the band gap. Unlike the case of 3D halide perovskites, not all types of octahedral tilts and rotations open the band gap; the contribution of Pb and X site distortions to the total distortion amplitude is shown to be decisive for predicting the effect of specific distortion modes on the band gap. Furthermore, we show that out of a total of 24 polar structures within $n=1$ and $n=2$, only four exhibit a pronounced Rashba-/Dresselhaus splitting. In particular, the ground state of $n=2$ is polar, but has negligible Rashba-/Dresselhaus splitting because the polar distortion modes are dominated by an alternating displacement of the Cs ions which do not contribute electronically to the band edges. Finally, to isolate the effect of interlayer stacking in RP halide perovskites, we briefly discuss results for model 2D systems consisting of free-standing halide perovskite mono- and bilayers, respectively. We find that the energy landscape and available band gap ranges are similar to the RP perovskites. However, the lack of interlayer stacking leads to a much smaller number of possible space groups and a reduced number of polar space groups for the monolayer.

We start our discussion with a description of our methods and nomenclature for all RP halide perovskite structures with $n=1$ and $n=2$ in Section~\ref{sec:methods}. In Sections~\ref{sec:energies} and \ref{sec:bandgaps} we discuss the energy landscape and range of band gaps accessible via octahedral tilting distortions. For clarity we focus on materials with X=I and provide complete results for X=Br and X=Cl in the Appendices. We then focus on RP halide perovskites with non-centrosymmetric space groups and their macroscopic electronic polarization and Rashba-/Dresselhaus splitting in Section~\ref{sec:Rashba}. Section~\ref{sec:2D} is dedicated to a discussion of 2D mono- and bilayer halide perovskites and the energy landscape and electronic structure of these model systems. The final Section~\ref{sec:finalsection} summarizes our conclusions.

\section{Methods} \label{sec:methods}

\subsection{Computational Methods}
First-principles DFT calculations were performed using the Vienna Ab initio Simulation Package (VASP)~\cite{Kresse1996a, Kresse1996b}. We used a cutoff energy of 500\,eV for the expansion of Kohn-Sham orbitals in the plane wave basis, and projector augmented-wave pseudo-potentials~\cite{Kresse1999} with the following electronic configurations: Cs ($5s^2 5p^6 6s^1$), Pb ($6s^2 5d^{10} 6s^2$), I ($5s^2 5p^5$), Br ($4s^2 4p^5$), and Cl ($3s^2 3p^5$).. The exchange-correlation functional of Perdew, Burke, and Ernzerhofer (PBE)~\cite{Perdew1996} was used for all calculations, and spin-orbit-coupling (SOC) was included in the band structure calculations of Section~\ref{sec:Rashba}. PBE is known to overestimate lattice volumes ($\sim$ 4\%) and underestimate band gaps in halide perovskites. Since these are systematic errors and we are primarily interested in the trends of how octahedral tilting distortions affect the energy landscape, we performed all total energy calculations with the PBE functional and symmetry adapted $\mathbf k$-point grids using a $6 \times 6 \times 6$ mesh for cubic CsPbI$_3$ as reference. All structural relaxations were performed  until forces were smaller than 0.001\,eV/\AA{}. Symmetry mode analysis is performed with \textsc{isodistort} \cite{Campbell2006}. Macroscopic polarization is computed within the modern theory of polarization. Phonon dispersions are obtained within the harmonic approximation using the Phonopy code \cite{Togo2015}. Additionally, we performed band gap calculations for selected cases using Green's function-based many-body perturbation theory in the "one-shot" $G_0W_0$ approximation to corroborate the trends we found using DFT-PBE and with SOC included self-consistently. The $GW$ calculations are performed with the \textsc{BerkeleyGW}~\cite{Deslippe2012} code. The zeroth-order one-particle Green's function $G_0$ and the screened Coulomb interaction $W_0$ are obtained from PBE eigenfunctions and eigenvalues calculated with the \textsc{Quantum Espresso}~\cite{Giannozzi2009, Giannozzi2017} software package. These DFT-PBE calculations use a plane-wave cutoff energy of 60\,Ry, and norm-conserving fully-relativistic pseudopotentials~\cite{VanSetten2018}. $G_0W_0$ calculations use a polarizability cutoff of 8\,Ry, 1200\,bands and a half-shifted $\mathbf{k}$-point grid consisting of $4 \times 4 \times 4$ and $4 \times 4 \times 2$ points for structures with $Cmce$ and $I4/mmm$ symmetry, respectively.

\subsection{Structural Notation}
\begin{figure}[ht]
\centering
\includegraphics[width=0.5\linewidth]{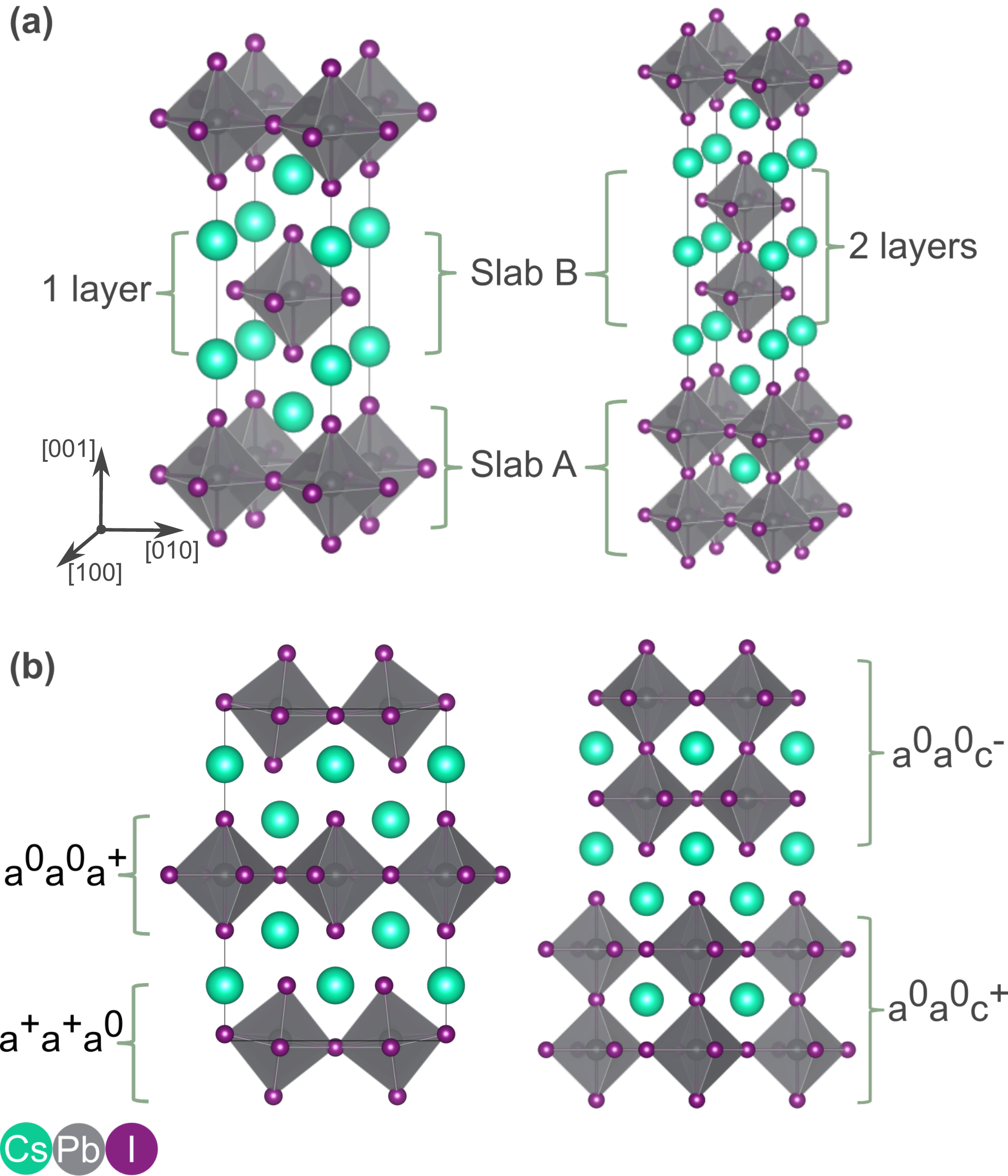}
\caption{\label{fig:structures}(a) Structure of $n=1$ \cspbiRpSingle{} (left) and $n=2$ \cspbiRpDouble{} (right) RP perovskites. Both structures consist of two perovskites slabs A and B. (b) Left: $n=1$ structure with tilt pattern \glazer{a}{+}{a}{+}{a}{0} in A and \glazer{a}{0}{a}{0}{a}{-} in B. Full tilt pattern: (\glazer{a}{+}{a}{+}{a}{0}/\glazer{a}{0}{a}{0}{a}{-},111). Right: $n=2$ structure with in- and out-of-phase rotation in $c$-direction in A and B. Full tilt pattern: (\glazer{a}{0}{a}{0}{a}{+}/\glazer{a}{0}{a}{0}{a}{-},111).}
\end{figure}

The all-inorganic RP structure with formula A$_{n+1}$B$_n$X$_{3n+1}$, shown in Figure~\ref{fig:structures}(a) can be described as a series of alternating ABX$_3$ perovskite and AX rock-salt layers. The rock-salt layer breaks the connectivity of the BX$_6$ octahedra in $c$-direction (defined as the out-of-plane direction [001]), creating two quasi-2D perovskite slabs A and B, which are shifted by ($a/2$, $a/2$, 0) in the $ab$-plane against each other ($a$ is the in-plane lattice parameter). The number of layers per slab is denoted by $n$. The undistorted reference structures for $n=1$ and $n=2$ with space group $I4/mmm$ are shown in Figure~\ref{fig:structures}(a). Starting from the reference structures for $n=1$ and $n=2$, we constructed unique space groups arising from all possible octahedral tilt patterns. The magnitude of the initial tilting angles in the input structures for geometry optimization was set to $\sim$ 6\textdegree.

To classify the octahedral tilt pattern of this perovskite family, we adapted the Glazer notation to account for the additional degrees of freedom that arise due to the broken octahedral connectivity in $c$-direction as shown in Figure~\ref{fig:structures}(b). Our notation consists of two tilt patterns, e.g. $a^-a^0a^0/a^0a^+a^+$, corresponding to the tilt patterns of slab A and slab B, respectively. Note that for $n=1$ RP perovskites, one cannot distinguish between in- and out-of-phase rotation along the $c$-direction, as each slab consists of only one perovskite layer. In this case, we use the "$+$"\,sign to indicate the presence of a rotation along $c$. Finally, due to the separation of the slabs, there is no unique definition for the direction of rotation of adjacent octahedra of different slabs \cite{li_suppressing_2020}. We therefore used three additional values $\pm 1$, denoting the relative rotation of octahedra in slabs A and B around the three cartesian axes, where $+1$ stands for in-phase and $-1$ for out-of-phase rotations (see Figure~\ref{fig:structures}b).

Applying all different combinations of tilt patterns and the different relative rotations of slab A and B to the undistorted $I4/mmm$ structure leads to 38 and 55 unique space groups for $n=1$ and $n=2$, respectively. From these, 9 structures of $n=1$ and 15 structures of $n=2$ are polar. Additionally, for both $n=1$ and $n=2$ we include structures with pure lead displacements: $Fmm2$, $Imm2$ and $I4mm$, where the Pb ions are displaced along the in-plane out-of-bond $a-/b-$, in-plane in-bond $a-$ and out-of-plane $c-$ direction, respectively. The lists of all space groups for $n=1$ and $n=2$ are provided in Tables~\ref{tab:n1Energies} and \ref{tab:n2Energies}, respectively (polar space groups are highlighted in green). Corresponding tables for the $n=1$ and $n=2$ bromides and chlorides are shown in~\ref{sec:appendix1}. Note that the tilt patterns reported in these Tables denote initial structures used as starting points, without structural relaxation. The space group symmetry only rarely changed throughout relaxation. For the iodides, this only happened for the \cspbiRpSingle{} structure with initial space group $Pm$ which turned into $Pmc2_1$. For the bromides and chlorides, symmetry changes during relaxation were observed in several cases, owing to the smaller spread in total energies as compared to the iodides (see Section~\ref{sec:energies}).

\begin{table}[ht]
  \centering
  \caption{Initial tilt pattern, energy gain $\Delta$E, size and position of the band gap for \cspbiRpSingle{}. All values were calculated using PBE, without SOC. Polar space groups are marked in green.}
  \resizebox*{0.6\textwidth}{!}{
\begin{tabular}{rllllrr}
    \br
    \multicolumn{2}{l}{space group}        &\multicolumn{2}{c}{tilt pattern} & $\mathbf k$-point & \multicolumn{1}{l}{$\Delta$E (meV)} & \multicolumn{1}{l}{band gap (eV)} \\ \mr
    
   \polColorRow{} 1     & P1    & a$^-$a$^+$a$^0$/a$^-$a$^+$a$^+$ & 111   & $\Gamma$ & -61.76 & 2.00 \\
    2     & P\minus{1}   & a$^0$b$^-$b$^+$/a$^0$b$^-$b$^+$ & 11-1  & Z     & -62.64 & 2.00 \\
   \polColorRow{} 4     & P2$_1$ & a$^+$a$^-$a$^+$/a$^+$a$^-$a$^+$ & 111   & $\Gamma$ & -60.62 & 2.00 \\
   \polColorRow{} 5     & C2    & a$^-$a$^+$a$^+$/a$^+$a$^-$a$^+$ & 111   & $\Gamma$ & -46.32 & 2.02 \\
   \polColorRow{} 6     & Pm    & a$^-$a$^+$a$^+$/a$^0$a$^+$a$^0$ & 111   & $\Gamma$ & -67.41 & 2.02 \\
   \polColorRow{} 7     & Pc    & a$^+$b$^+$b$^+$/a$^+$b$^+$b$^+$ & 11-1  & $\Gamma$ & -31.92 & 2.09 \\
   \polColorRow{} 9     & Cc    & a$^-$a$^+$a$^+$/a$^+$a$^-$a$^+$ & 1-11  & $\Gamma$ & -66.46 & 2.03 \\
    10    & P2/m  & a$^-$a$^0$a$^+$/a$^0$a$^+$a$^0$ & 111   & $\Gamma$ & -68.04 & 2.02 \\
    11    & P2$_1$/m & a$^-$a$^0$a$^0$/a$^0$a$^+$a$^+$ & 111   & $\Gamma$ & -68.24 & 2.02 \\
    12    & C2/m  & a$^-$a$^-$a$^0$/a$^+$a$^+$a$^0$ & 111   & $\Gamma$ & -58.48 & 2.02 \\
    13    & P2/c  & a$^-$b$^-$c$^0$/a$^-$b$^-$c$^0$ & 1-11  & $\Gamma$ & -62.99 & 1.99 \\
    14    & P2$_1$/c & a$^0$b$^-$b$^+$/a$^0$b$^-$b$^+$ & 111   & Y     & -60.34 & 2.16 \\
    15    & C2/c  & a$^0$a$^-$c$^+$/a$^0$a$^-$c$^+$ & 111   & $\Gamma$ & -41.73 & 2.13 \\
    18    & P2$_1$2$_1$2 & a$^-$a$^+$a$^0$/a$^+$a$^0$a$^0$ & 111   & $\Gamma$ & -64.62 & 2.04 \\
   \polColorRow{} 26    & Pmc2$_1$ & a$^-$a$^+$a$^0$/a$^0$a$^+$a$^0$ & 111   & $\Gamma$ & -67.08 & 2.02 \\
   \polColorRow{} 31    & Pmn2$_1$ & a$^0$a$^+$a$^0$/a$^0$a$^+$a$^+$ & 111   & $\Gamma$ & -25.82 & 2.05 \\
   \polColorRow{} 41    & Aea2  & a$^+$a$^+$a$^+$/a$^+$a$^+$a$^+$ & 11-1  & $\Gamma$ & -12.35 & 2.02 \\
    51    & Pmma  & a$^-$a$^0$a$^0$/a$^0$a$^+$a$^0$ & 111   & $\Gamma$ & -68.08 & 2.02 \\
    53    & Pmna  & a$^-$a$^-$a$^0$/a$^0$a$^0$a$^0$ & 111   & $\Gamma$ & -59.77 & 2.02 \\
    55    & Pbam  & a$^0$a$^0$a$^+$/a$^0$a$^0$c$^+$ & 111   & $\Gamma$ & -1.28 & 2.12 \\
    56    & Pccn  & a$^-$a$^-$c$^+$/a$^-$a$^-$c$^+$ & 111   & $\Gamma$ & -44.54 & 2.10 \\
    57    & Pbcm  & a$^-$a$^0$a$^0$/a$^+$a$^-$a$^0$ & 111   & $\Gamma$ & -45.71 & 1.99 \\
    59    & Pmmn  & a$^0$a$^0$a$^0$/a$^0$a$^+$a$^+$ & 111   & $\Gamma$ & 1.02  & 1.90 \\
    61    & Pbca  & a$^-$a$^-$c$^+$/a$^-$a$^-$c$^+$ & -1-1-1 & $\Gamma$ & -63.67 & 2.13 \\
    62    & Pnma  & a$^0$a$^+$c$^0$/a$^0$a$^+$c$^0$ & 111   & X     & -24.04 & 2.05 \\
    64    & Cmce  & a$^0$a$^0$c$^+$/a$^0$a$^0$c$^+$ & 111   & $\Gamma$ & -1.30 & 2.18 \\
    66    & Cccm  & a$^-$a$^-$c$^0$/a$^-$a$^-$c$^0$ & 111   & $\Gamma$ & -41.24 & 1.97 \\
    67    & Cmme  & a$^-$a$^0$a$^0$/a$^0$a$^0$a$^0$ & 111   & $\Gamma$ & -69.30 & 2.01 \\
    85    & P4/n  & a$^0$a$^0$a$^+$/a$^+$a$^+$a$^0$ & 111   & $\Gamma$ & 1.42  & 1.99 \\
    86    & P4$_2$/n & a$^-$a$^-$c$^0$/a$^-$a$^-$c$^0$ & 1-11  & $\Gamma$ & -69.26 & 2.02 \\
    94    & P4$_2$2$_1$2 & a$^-$a$^+$a$^0$/a$^+$a$^-$a$^0$ & 111   & $\Gamma$ & -45.48 & 2.03 \\
    114   & P\minus{4}2$_1$c & a$^-$a$^+$a$^0$/a$^+$a$^-$a$^0$ & 1-11  & $\Gamma$ & -66.39 & 2.04 \\
    127   & P4/mbm & a$^0$a$^0$a$^0$/a$^0$a$^0$a$^+$ & 111   & $\Gamma$ & 2.92  & 1.91 \\
    129   & P4/nmm & a$^0$a$^0$a$^0$/a$^+$a$^+$a$^0$ & 111   & $\Gamma$ & 1.32  & 1.91 \\
    134   & P4$_2$/nnm & a$^-$a$^0$a$^0$/a$^0$a$^-$a$^0$ & 111   & $\Gamma$ & -46.82 & 1.99 \\
    137   & P4$_2$/nmc & a$^0$a$^+$a$^0$/a$^+$a$^0$a$^0$ & 111   & $\Gamma$ & -2.60 & 1.97 \\
    138   & P4$_2$/ncm & a$^-$a$^0$a$^0$/a$^0$a$^-$a$^0$ & 1-11  & $\Gamma$ & -69.23 & 2.02 \\
    139   & I4/mmm & a$^0$a$^0$a$^0$/a$^0$a$^0$a$^0$ & 111   & $\Gamma$      & 0.00  & 1.90 \\
    \polColorRow{} 42    & Fmm2  & a$^0$a$^0$a$^0$/a$^0$a$^0$a$^0$ & 111 &  M    &  -0.69 &  1.93 \\
    \polColorRow{} 44    & Imm2  & a$^0$a$^0$a$^0$/a$^0$a$^0$a$^0$ & 111 & M     & -0.18 & 1.93 \\
   \polColorRow{} 107   & I4mm  & a$^0$a$^0$a$^0$/a$^0$a$^0$a$^0$ & 111 & $\Gamma$      & 0.20  & 1.90 \\\br
    \end{tabular}
      }
  \label{tab:n1Energies}%
\end{table}%

\begin{table}[ht]
  \centering
  \caption{Initial tilt pattern, energy gain $\Delta$E, size and position of the band gap for \cspbiRpDouble{}. All values were calculated using PBE, without SOC. Polar space groups are marked in green.}
  \resizebox*{0.6\textwidth}{!}{
\begin{tabular}{rllllrr}
    \br
    \multicolumn{2}{l}{space group}        &\multicolumn{2}{c}{tilt pattern} & $\mathbf k$-point & \multicolumn{1}{l}{$\Delta$E (meV)} & \multicolumn{1}{l}{band gap (eV)} \\\mr
    \polColorRow{} 1     & P1    & a$^-$a$^-$a$^-$/a$^-$a$^-$a$^+$ & 1-11  & $\Gamma$ & -108.80 & 1.98 \\
    2     & P\minus{1}   & a$^0$b$^-$b$^-$/a$^0$b$^-$b$^-$ & 111   & Z     & -124.07 & 1.86 \\
    \polColorRow{} 3     & P2    & a$^-$a$^0$a$^-$/a$^0$a$^+$a$^+$ & 111   & $\Gamma$ & -118.86 & 1.90 \\
    \polColorRow{} 4     & P21   & a$^+$a$^-$a$^-$/a$^+$a$^-$a$^-$ & 11-1  & $\Gamma$ & -112.20 & 1.96 \\
    \polColorRow{} 5     & C2    & a$^-$a$^-$a$^-$/a$^+$a$^+$a$^-$ & 111   & $\Gamma$ & -91.39 & 1.94 \\
    \polColorRow{} 6     & Pm    & a$^+$b$^+$b$^+$/a$^+$b$^+$b$^+$ & 11-1  & $\Gamma$ & -90.84 & 1.94 \\
    \polColorRow{} 7     & Pc    & a$^-$a$^-$a$^-$/a$^-$a$^-$a$^+$ & 111   & $\Gamma$ & -97.30 & 1.96 \\
    \polColorRow{} 8     & Cm    & a$^0$b$^-$b$^+$/a$^0$b$^-$b$^+$ & 111   & $\Gamma$ & -125.45 & 1.86 \\
    10    & P2/m  & a$^-$b$^-$c$^0$/a$^-$b$^-$c$^0$ & 1-11  & $\Gamma$ & -125.47 & 1.87 \\
    11    & P21/m & a$^+$a$^-$c$^0$/a$^+$a$^-$c$^0$ & 111   & $\Gamma$ & -122.71 & 1.87 \\
    12    & C2/m  & a$^0$a$^-$c$^0$/a$^0$a$^-$c$^0$ & 111   & $\Gamma$ & -125.42 & 1.86 \\
    13    & P2/c  & a$^-$a$^-$a$^-$/a$^-$a$^-$a$^-$ & 111   & $\Gamma$ & -82.81 & 1.96 \\
    14    & P21/c & a$^0$b$^-$b$^-$/a$^0$b$^-$b$^-$ & 11-1  & $\Gamma$ & -124.98 & 1.86 \\
    15    & C2/c  & a$^-$a$^-$a$^-$/a$^-$a$^-$a$^-$ & -1-1-1 & $\Gamma$ & -101.22 & 1.99 \\
    16    & P222  & a$^0$a$^0$a$^-$/a$^0$a$^+$a$^+$ & 111   & $\Gamma$ & -36.20 & 1.84 \\
    17    & P222$_1$ & a$^0$a$^0$a$^+$/a$^0$a$^+$a$^-$ & 111   & $\Gamma$ & -41.75 & 1.89 \\
    20    & C2221 & a$^+$a$^+$a$^-$/a$^+$a$^+$a$^-$ & 111   & $\Gamma$ & -72.96 & 1.97 \\
    21    & C222  & a$^0$a$^0$a$^-$/a$^+$a$^+$a$^-$ & 111   & $\Gamma$ & -34.87 & 1.95 \\
    \polColorRow{} 25    & Pmm2  & a$^-$a$^0$a$^0$/a$^0$a$^+$a$^+$ & 111   & $\Gamma$ & -119.15 & 1.90 \\
    \polColorRow{} 26    & Pmc2$_1$ & a$^-$a$^-$a$^0$/a$^-$a$^-$a$^+$ & 111   & $\Gamma$ & -90.55 & 1.92 \\
    \polColorRow{} 28    & Pma2  & a$^-$a$^0$a$^+$/a$^0$a$^-$a$^+$ & 111   & $\Gamma$ & -99.78 & 1.91 \\
    \polColorRow{} 31    & Pmn2$_1$ & a$^-$a$^0$a$^+$/a$^0$a$^-$a$^+$ & 1-11  & $\Gamma$ & -120.88 & 1.91 \\
    \polColorRow{} 36    & Cmc2$_1$ & a$^-$a$^-$a$^+$/a$^-$a$^-$a$^+$ & -1-1-1 & $\Gamma$ & -136.64 & 2.01 \\
    \polColorRow{} 38    & Amm2  & a$^-$a$^-$a$^0$/a$^+$a$^+$a$^0$ & 111   & $\Gamma$ & -95.80 & 1.95 \\
    \polColorRow{} 39    & Aem2  & a$^-$a$^-$a$^+$/a$^-$a$^-$a$^+$ & 111   & $\Gamma$ & -117.58 & 2.00 \\
    \polColorRow{} 40    & Ama2  & a$^+$a$^+$a$^+$/a$^+$a$^+$a$^+$ & 11-1  & $\Gamma$ & -96.30 & 1.94 \\
    47    & Pmmm  & a$^-$a$^0$a$^0$/a$^0$a$^+$a$^0$ & 111   & $\Gamma$ & -73.57 & 1.80 \\
    49    & Pccm  & a$^0$a$^0$a$^-$/a$^0$a$^+$a$^0$ & 111   & $\Gamma$ & -36.13 & 1.83 \\
    50    & Pban  & a$^0$a$^0$a$^-$/a$^0$a$^0$c$^-$ & 111   & $\Gamma$ & -36.46 & 2.06 \\
    51    & Pmma  & a$^-$a$^-$a$^0$/a$^0$a$^0$a$^0$ & 111   & $\Gamma$ & -98.02 & 1.94 \\
    54    & Pcca  & a$^-$a$^-$a$^-$/a$^-$a$^-$a$^-$ & 11-1  & $\Gamma$ & -86.54 & 1.94 \\
    55    & Pbam  & a$^0$a$^0$a$^+$/a$^0$a$^0$c$^+$ & 111   & $\Gamma$ & -35.25 & 2.03 \\
    57    & Pbcm  & a$^-$a$^-$a$^+$/a$^-$a$^-$a$^+$ & 11-1  & $\Gamma$ & -112.59 & 2.01 \\
    59    & Pmmn  & a$^0$a$^+$c$^0$/a$^0$a$^+$c$^0$ & 111   & X     & -70.29 & 1.88 \\
    60    & Pbcn  & a$^-$a$^-$a$^-$/a$^-$a$^-$a$^-$ & -1-11 & $\Gamma$ & -105.40 & 1.96 \\
    62    & Pnma  & a$^-$a$^-$a$^+$/a$^-$a$^-$a$^+$ & -1-11 & $\Gamma$ & -131.45 & 2.02 \\
    63    & Cmcm  & a$^-$a$^-$c$^0$/a$^-$a$^-$c$^0$ & -1-11 & $\Gamma$ & -98.04 & 1.95 \\
    64    & Cmce  & a$^0$a$^0$c$^+$/a$^0$a$^0$c$^+$ & 111   & $\Gamma$ & -35.21 & 2.06 \\
    65    & Cmmm  & a$^-$a$^0$a$^0$/a$^0$a$^0$a$^0$ & 111   & $\Gamma$ & -121.14 & 1.90 \\
    66    & Cccm  & a$^-$a$^+$a$^0$/a$^+$a$^-$a$^0$ & 111   & $\Gamma$ & -97.47 & 1.94 \\
    67    & Cmme  & a$^-$a$^-$c$^0$/a$^-$a$^-$c$^0$ & 111   & $\Gamma$ & -79.37 & 1.92 \\
    68    & Ccce  & a$^0$a$^0$c$^-$/a$^0$a$^0$c$^-$ & 111   & $\Gamma$ & -36.43 & 2.08 \\
    81    & P\minus{4}   & a$^0$a$^0$a$^+$/a$^+$a$^+$a$^-$ & 111   & $\Gamma$ & -38.60 & 1.96 \\
    83    & P4/m  & a$^0$a$^0$a$^+$/a$^+$a$^+$a$^0$ & 111   & $\Gamma$ & -30.37 & 1.89 \\
    84    & P42/m & a$^-$a$^-$c$^0$/a$^-$a$^-$c$^0$ & 1-11  & $\Gamma$ & -120.57 & 1.90 \\
    89    & P422  & a$^0$a$^0$a$^-$/a$^+$a$^+$a$^0$ & 111   & $\Gamma$ & -35.04 & 1.91 \\
    115   & P\minus{4}m2 & a$^0$a$^0$a$^0$/a$^+$a$^+$a$^-$ & 111   & $\Gamma$ & -13.82 & 1.79 \\
    117   & P\minus{4}b2 & a$^0$a$^0$a$^-$/a$^0$a$^0$a$^+$ & 111   & $\Gamma$ & -35.90 & 2.05 \\
    123   & P4/mmm & a$^0$a$^0$a$^0$/a$^+$a$^+$a$^0$ & 111   & $\Gamma$ & -13.99 & 1.79 \\
    125   & P4/nbm & a$^0$a$^0$a$^-$/a$^0$a$^0$a$^0$ & 111   & $\Gamma$ & -3.57 & 1.78 \\
    127   & P4/mbm & a$^0$a$^0$a$^0$/a$^0$a$^0$a$^+$ & 111   & $\Gamma$ & -2.92 & 1.77 \\
    131   & P4$_2$/mmc & a$^0$a$^+$a$^0$/a$^+$a$^0$a$^0$ & 111   & $\Gamma$ & -40.08 & 1.87 \\
    132   & P4$_2$/mcm & a$^-$a$^0$a$^0$/a$^0$a$^-$a$^0$ & 111   & $\Gamma$ & -100.35 & 1.90 \\
    136   & P42/mnm & a$^-$a$^0$a$^0$/a$^0$a$^-$a$^0$ & 1-11  & $\Gamma$ & -120.82 & 1.90 \\
    139   & I4/mmm & a$^0$a$^0$c$^0$/a$^0$a$^0$a$^0$ & 111   & $\Gamma$      & 0.00  & 1.78 \\
    \polColorRow{} 42    & Fmm2  & a$^0$a$^0$c$^0$/a$^0$a$^0$a$^0$ & 111 &  M    & -0.42 &  1.80 \\
    \polColorRow{} 44    & Imm2  & a$^0$a$^0$c$^0$/a$^0$a$^0$a$^0$ & 111 & M     & -0.20 & 1.80 \\
    \polColorRow{} 107   & I4mm  & a$^0$a$^0$c$^0$/a$^0$a$^0$a$^0$ & 111 & $\Gamma$      & -0.45 & 1.78 \\\br
    \end{tabular}
  \label{tab:n2Energies}%
  }
\end{table}%

\section{Energy Landscape of RP Halide Perovskites} \label{sec:energies}
We start by investigating the energy landscape of all possible \cspbiRpSingle{} and \cspbiRpDouble{} structures. The energy gain with respect to the $I4/mmm$ parent structure is listed in Tables~\ref{tab:n1Energies} and \ref{tab:n2Energies}, respectively. We first note that similar to 3D ABX$_3$ perovskites, displacive distortions resulting in space groups $Fmm2$, $Imm2$ and $I4mm$ are energetically unfavorable, whereas octahedral rotations and tilts lead to a lowering of the energy. For $n=1$ RP perovskites, the overall energy landscape is similar for all three halogen anions. The ground state of \cspbiRpSingle{} has $Cmme$ symmetry with an energy gain of -69.3\,meV. For \cspbbrRpSingle{} and \cspbclRpSingle{} the ground state has $P4_2/ncm$ symmetry with energy gains of -56.6\,meV and -51.2\,meV, respectively. All these structures have a tilt pattern of $a^-a^0a^0/a^0a^-a^0$, but differ in the magnitudes of the rotation angles. Furthermore, in all cases there are several other structures that are energetically very close to the ground state with energy differences smaller than $\sim$10\,meV. For \cspbiRpSingle{}, the lowest-energy polar structure has space group $Pmc2_1$ and is only 2\,meV higher in energy than the ground state. For \cspbbrRpSingle{} and \cspbclRpSingle{}, the polar space groups $Cc$ (X=Br) and $Pmc2_1$ (X=Cl) are within $\sim$1\,meV of the ground state. Contrasting this finding to the 3D perovskites CsPbI$_3$, CsPbBr$_3$, and CsPbCl$_3$, for which the lowest-energy polar structure with $R3c$ symmetry is found 43\,meV, 29\,meV, and 21\,meV above the non-polar ground state, suggests that polar space groups should be more easily accessible in RP perovskites than in their 3D congeners \cite{Leppert2016}.
\begin{table}[ht]
\centering
  \caption{Selected modes representing different tilt patterns and distortions in Cs$_2$PbI$_4$ ($n=1$) and Cs$_3$Pb$_2$I$_7$ ($n=2$) RP perovskites.}
\begin{tabular}{llll}
\br
\multicolumn{2}{l}{$n=1$}          & \multicolumn{2}{l}{$n=2$}         \\ %\mr
mode         & distortion pattern                   & mode         & distortion pattern                  \\ \mr
$X_2^+$      & $a^0a^0a^+$                  & $X_2^+$      & $a^0a^0a^+$                 \\
             &                              & $X_1^-$      & $a^0a^0a^-$                 \\    
$X_3^+$      & \multirow{2}{*}{$a^-a^-a^0$} & $X_3^-$      & \multirow{2}{*}{$a^-a^-a^0$}\\
$X_4^+$      &                              & $X_4^-$      &                              \\
$\Sigma_3$   & $a^+a^0a^0$                  & $\Sigma_1$   &   $a^+a^0a^0$                 \\
$\Gamma_3^-$ & displacement along $c$       & $\Gamma_3^-$ & displacement along $c$        \\
$\Gamma_5^-$ & displacement along $a$, $b$  & $\Gamma_5^-$ & displacement along $a$, $b$   \\\br
\end{tabular}
\label{tab:distortions}
\end{table}

For $n=2$, the ground state is the polar $Cmc2_1$ structure with energy gains of -136.6\,meV (\cspbiRpDouble{}), -102.2\,meV (\cspbbrRpDouble{}), and -90.4\,meV (\cspbclRpDouble{}) with respect to the reference structure $I4/mmm$. In contrast to $n=1$, the ground state is distinct and $\sim$5\,meV lower in energy than all other structures. Note that the tilt pattern of $Cmc2_1$ ($a^-a^-a^0/a^-a^-a^+$), corresponds to the tilt pattern of the low-energy $Pnma$ phase of the 3D ABX$_3$ perovskites, which is, however, not polar. This finding is in line with the observation that $Cmc2_1$ is a common ground state in perovskite oxides \cite{li_suppressing_2020, zhang_strain-induced_2017} and other RP perovskites \cite{wang_first-principles_2021, wang_ruddlesdenpopper_2016}. For the all-inorganic RP halide perovskite Cs$_3$Sn$_2$X$_7$ (X= I, Br), the ground state was found to be a structure with $P4_2/mnm$ symmetry. However, a transition to $Cmc2_1$ was observed due to the appearance of an in-phase rotation under compressive stress \cite{li_suppressing_2020}. Since Sn has a smaller ionic radius than Pb, the tolerance factor of \cspbiRpDouble{} is smaller than that of Cs$_3$Sn$_2$I$_7$, and the rotation required for the polar phase is stable even without strain \cite{travis_application_2016}. Finally, we note that all 15 polar structures of $n=2$ that arise due to octahedral tilting are within 46\,meV (\cspbiRpDouble{}), 34\,meV (\cspbbrRpDouble{}), and 30\,meV (\cspbclRpDouble{}) of the energy of the ground state structure, whereas the $I4mm$ and $Imm2$ structures are energetically unfavorable.

\begin{figure}[ht]
\centering
\includegraphics[width=0.5\linewidth]{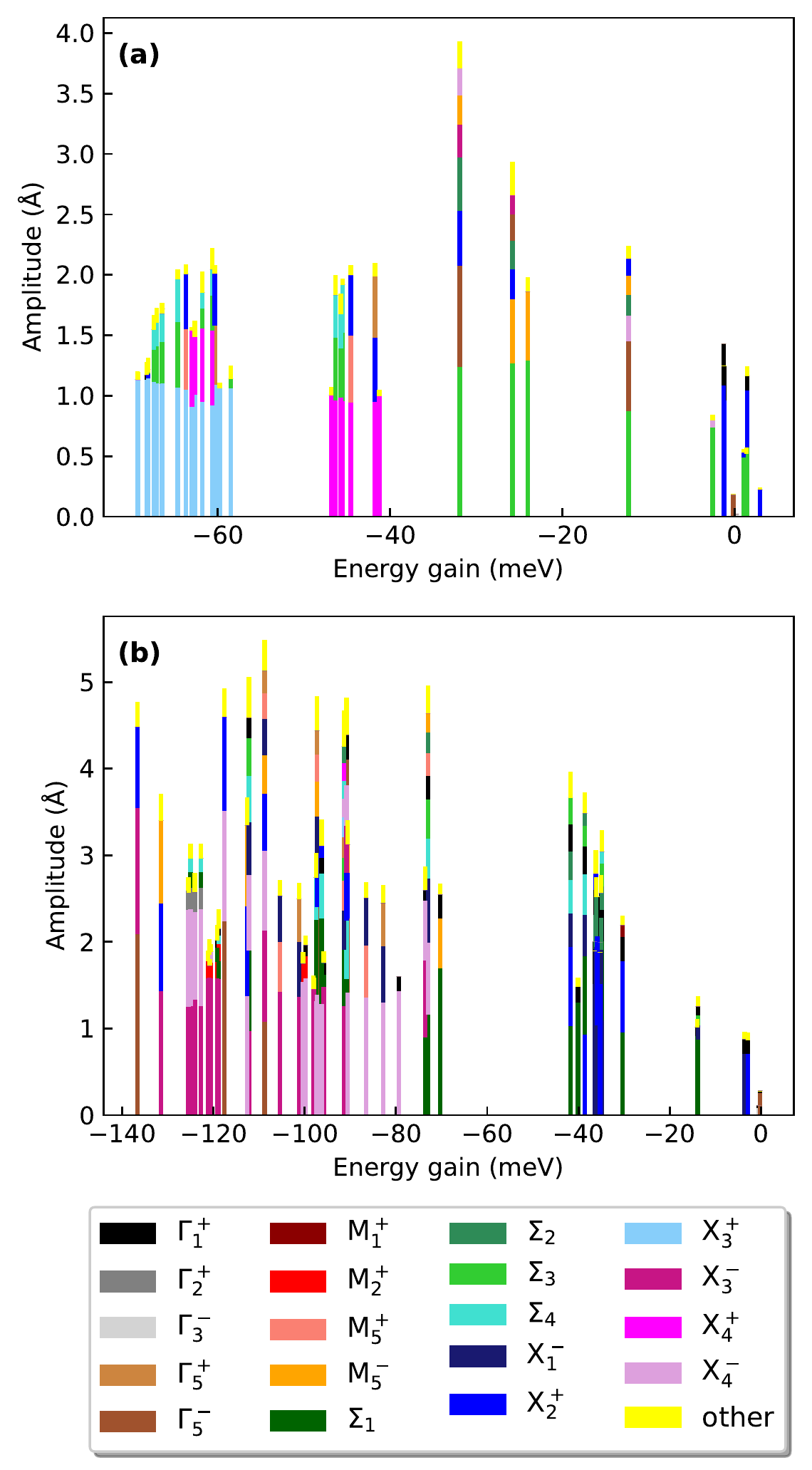}
\caption{\label{fig:energies} Mode decomposition for all polar structures within (a) $n=1$ \cspbiRpSingle{} and (b) $n=2$ \cspbiRpDouble{} RP structures, showing the amplitude of each distortion mode within the distorted structure as a function of energy gain with respect to the $I4/mmm$ parent structure. Each bar represents one structure and shows the total mode amplitude and its individual contributions. The modes are sorted by amplitude with the dominant one at the bottom. For clarity, we only included modes accounting for 90\% of the total amplitude. Color code is included at the bottom of the figure.}
\end{figure}  

To gain a better understanding of how octahedral rotations lower the energy of RP halide perovskites, we proceeded by decomposing every structure into distortion modes. An overview of selected modes and the distortions they correspond to is shown in Table~\ref{tab:distortions}. For example, $\Sigma_1$ and $\Sigma_3$ modes in $n=1$ and $n=2$, respectively, correspond to in-phase rotations along the in-plane $a$- and $b$-directions, whereas $X_2^+$ and $X_1^-$ describe the in- and out-of-phase rotation along the out-of-plane c-axis. The modes $X_3^\pm$ and $X_4^\pm$ correspond to equivalent tilt patterns but with different relative rotation between slabs A and B. Depending on which ions are affected by the mode, the tilt pattern occurs only in one or in both slabs. More complex tilt patterns like $a^-b^+b^+$ are a combination of these modes. Finally, the $\Gamma_3^-$ and $\Gamma_5^-$ modes describe polar displacements along the in-plane $a/b$ or out-of-plane $c$-directions, respectively. Phonon dispersions (shown in Figure~\ref{fig:phonon_bands_n=1}) for $n=1$ reveal symmetry breaking instabilities along several symmetry lines in the Brillouin zone, including the $\Gamma-X$ segment, dominated by octahedral tilts comprising I atom displacements. Similar results are expected for $n=2$, where the additional inorganic layer increases the degree of cooperativity of octahedral tilts along the out-of-plane c-direction. 

In Figures~\ref{fig:energies}(a) and (b), we show the mode decomposition of every \cspbiRpSingle{} and \cspbiRpDouble{} structure as a function of the energy gain with respect to the $I4/mmm$ parent compound. Each structure is composed of several distortion modes which we sorted such that the dominant mode with the largest amplitude is at the bottom of each bar. We normalize these amplitudes with respect to the $I4/mmm$ parent cell so that amplitudes of different space groups of the same layer number $n$ are comparable. For $n=1$ in Figure~\ref{fig:energies}(a), we observe a clear clustering of structures with the same dominant distortion mode. All of the lowest energy structures are dominated by an $X_3^+$ mode, followed by a group of structures dominated by an $X_4^+$ mode, and a group of structures with dominant $\Sigma_3$ mode. Within each cluster, structures can further be sorted based on their second-largest mode, in particular in the $X_3^+$-dominated group, where structures with secondary $\Sigma_3$ distortion are lower in energy than those with $X_4^+$ as the second largest mode.
  
For $n=2$ in Figure~\ref{fig:energies}(b), most of the dominant modes (specifically $X_3^-$, $X_4^-$, $\Sigma_1$, $X_2^+$, $X_1^-$) correspond to octahedral rotations. The large $\Gamma_5^-$ mode in the ground state structure with $Cmc2_1$ symmetry and the $Aem2$ structure at $\sim$-120\,meV originates from an alternating shift of the Cs ions that breaks the centrosymmetry of these structures \cite{mulder_turning_2013, wang_ruddlesdenpopper_2016}. In comparison to $n=1$, the grouping of structures with similar dominant distortion modes is less distinct. Furthermore, unlike for $n=1$ where in particular the low-energy structures are dominated by two to three distortion modes, for $n=2$ multiple modes with similar amplitudes contribute to the overall distortion.

Overall, we find that for both \cspbiRpSingle{} and \cspbiRpDouble{}, structures with a dominant mode corresponding to out-of-phase tilts in $a$/$b$-direction ($X_3^\pm$, $X_4^\pm$) are energetically most favorable, followed by in-phase tilts in $a$/$b$-direction ($\Sigma_3$, $\Sigma_1$). Rotation around the $c$-axis ($X_2^+$,$X_1^-$) is mainly found in energetically higher structures.

\section{Tuning of Band Gaps through Octahedral Rotations} \label{sec:bandgaps}
Octahedral tilts are well-known to strongly affect the band gaps of 3D ABX$_3$ perovskites \cite{Filip2014, Wiktor2017a}. In these materials, octahedral tilts open the band gap because they lead to a stabilization of both the anti-bonding VBM and CBM with a larger energy reduction of the VBM due to the bond-length sensitivity of the Pb-$6s$ lone pair that contributes to the VBM \cite{Zhao2021}. Here, we will explore to which extent different octahedral tilt patterns affect the band gap of RP halide perovskites. Unless otherwise noted, our band gaps are obtained with DFT-PBE calculations, without including SOC effects. Previous work has shown an error cancellation in the magnitude of DFT-PBE band gaps, since the spin-orbit splitting of the CBM's Pb $p$ states lowers the band gap of Pb-based halide perovskite by $\sim$1\,eV, whereas the inherent DFT band gap underestimation with semilocal functionals such as PBE leads to an underestimation of the band gap of similar size. We believe that this approach is justified in our case since we are only interested in trends and not in accurate band gap predictions. Furthermore, we have confirmed the accuracy of our reported band gap ranges by performing additional G$_0$W$_0$@PBE calculations including SOC, which gives reliable band gaps for a wide range of halide perovskites \cite{Umari2013, Filip2014d, Filip2016, Leppert2019l}.

The DFT-PBE band gaps of all structures of \cspbiRpSingle{} and \cspbiRpDouble{} can be found in Tables~\ref{tab:n1Energies} and \ref{tab:n2Energies}, respectively. Results for the bromides and chlorides are listed in the \ref{sec:appendix1} and \ref{sec:appendix2}, and follow similar trends. The Tables also report the $\mathbf k$-point at which the band gap of these semiconductors is located. In most cases, the band gap is direct at the $\Gamma$ point, or an equivalent high-symmetry point for non primitive unit cells after band unfolding. The band gaps of the structures with the lowest and highest band gap, and of the ground state structure are reported in Table~\ref{tab:gaprange}. In both $n=1$ and $n=2$, the band gap difference between the lowest- and the highest band gap structure is $\sim$0.3\,eV, for all three halides. Our $G_0W_0$ calculations confirm the error cancellation between SOC and quasiparticle corrections observed before \cite{Filip2014d}, resulting in band gaps very close to the ones calculated with DFT-PBE. The ground state structure features a band gap right in between the range of accessible band gaps in all cases. Furthermore, the structure with the lowest band gap is the undistorted parent structure $I4/mmm$ only in the case of \cspbiRpSingle{}, raising the question how exactly octahedral rotations affect the band gap of these materials.

\begin{table}[ht]
\centering
\caption{DFT-PBE band gaps for the Cs$_2$PbI$_4$ ($n=1$) and Cs$_3$Pb$_2$I$_7$ ($n=2$) structures with the lowest and the highest band gap, and of the ground state structure. $G_0W_0$@PBE+SOC band gaps of \cspbiRpSingle{} with $I4/mmm$ and $Cmce$ symmetry are reported in parenthesis.}
\begin{tabular}{ccccc}
\br
X site              & \multicolumn{2}{c}{$n=1$} & \multicolumn{2}{c}{$n=2$} \\%\mr
                    & space group & E$_{gap}$ (eV) & space group & E$_{gap}$ (eV)         \\ \mr
\multirow{3}{*}{I}  & $I4/mmm$    & 1.90 (1.87)        & $P4/mbm$    & 1.77               \\
                    & $Cmce$      & 2.18  (2.12)       & $Ccce$      & 2.09               \\
                    & $Cmme$      & 2.01               & $Cmc2_1$    & 2.01               \\\mr
\multirow{3}{*}{Br} & $Pmmn$      & 2.27               & $P4/mbm$    & 2.10               \\
                    & $Cmce$      & 2.55               & $Ccce$      & 2.40               \\
                    & $P4_2/ncm$  & 2.38               & $Cmc2_1$    & 2.35               \\\mr
\multirow{3}{*}{Cl} & $Cccm$      & 2.68               & $P4/mbm$    & 2.52               \\
                    & $Cmce$      & 2.99               & $Ccce$      & 2.81               \\
                    & $P4_2/ncm$  & 2.84               & $Cmc2_1$    & 2.78               \\\br
\end{tabular}
\label{tab:gaprange}
\end{table}

\begin{figure}[ht]
\centering
\includegraphics[width=0.6\linewidth]{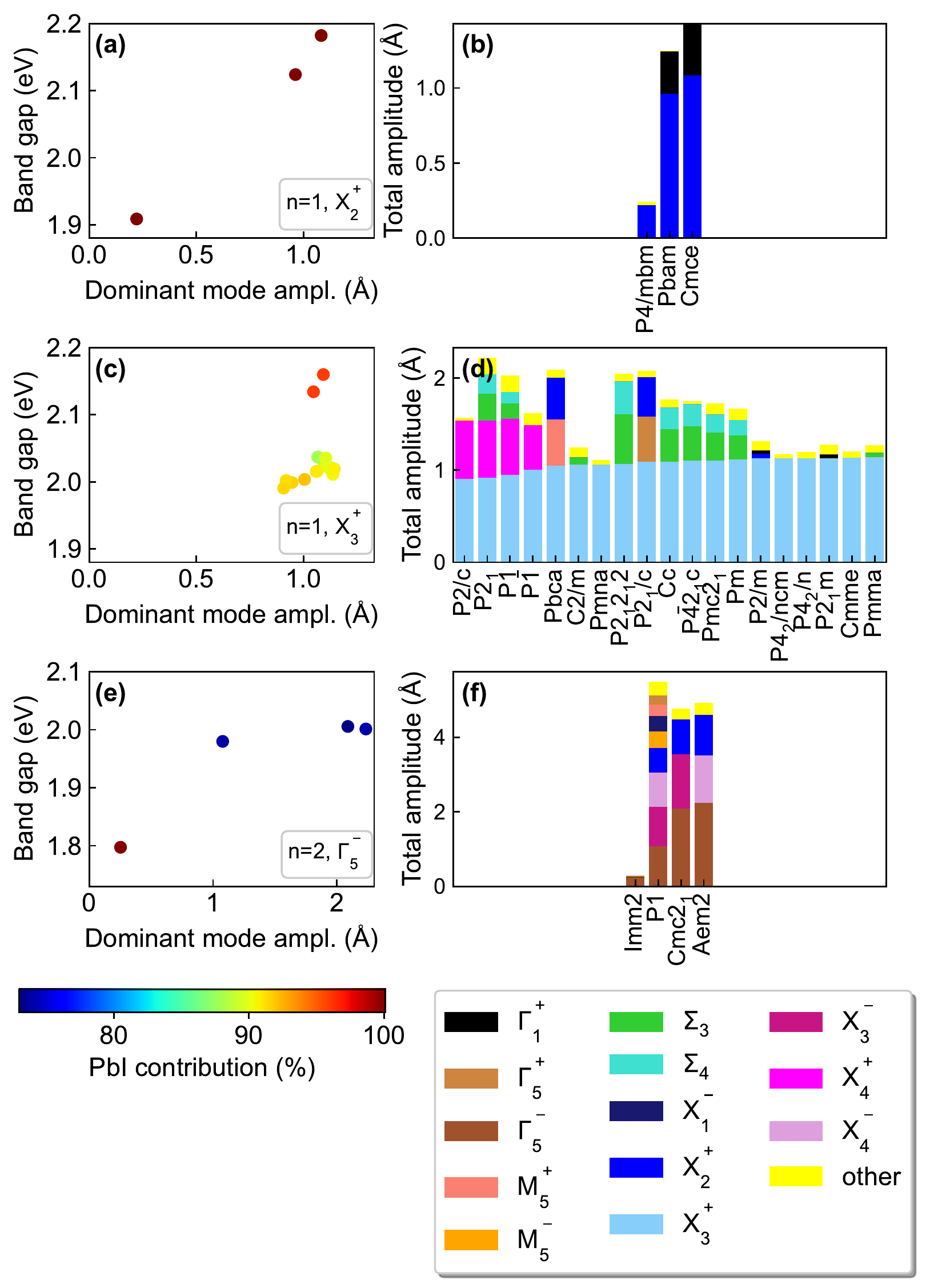}
\caption{\label{fig:gaps} (left column) Band gap size with respect to the amplitude of the dominant distortion mode. The color of the dot shows how much of the mode amplitude stems from the displacement of Pb- and I-ions, as represented on the color scale below the figure. (right column) Mode decomposition of the structures shown on the left side. The structures are sorted by increasing amplitude of the dominant mode and the color code is included at the bottom of the figure. (a,b) shows $n=1$ structures with a dominant $X_2^+$ mode, (c,d) $n=1$ structures with a dominant $X_3^+$ mode and (e,f) $n=2$ structures with a dominant $\Gamma_5^-$ mode.}
\end{figure}

We proceeded by analysing the band gap variation in terms of dominant symmetry-mode distortions, distinguishing three classes of structures. In Figure~\ref{fig:gaps}, we illustrate the most important trends for three representative cases: 
\begin{enumerate}
    \item First case: Structures for which the band gap increases with the amplitude of the dominant distortion mode (determined in the same way as in Section~\ref{sec:energies}) and depicted in Figures~\ref{fig:gaps}(a) and (b) for the case of \cspbiRpSingle{} structures with dominant $X_2^+$ mode.
    \item Second case: Structures with a large band gap variation that does not depend on the amplitude of the dominant mode, shown in Figures~\ref{fig:gaps}(c) and (d) for the case of \cspbiRpSingle{} structures in which the dominant distortion mode is $X_3^+$.
    \item Third case: Structures with no correlation between the amplitude of the dominant mode and the size of the band gap, shown in Figures~\ref{fig:gaps}(e) and (f) for the case of \cspbiRpDouble{} with dominant $\Gamma_5^-$ mode.
\end{enumerate}

It is important to note that considering all possible $n=1$ and $n=2$ structures, the correlation between band gap size and the amplitude of the dominant mode is weak. This is not surprising since the band gap is a macroscopic property of the material which is determined by an interplay of all modes. However, the picture barely changes when the second-largest mode is taken into account. These findings can be explained by analysing the amplitude of the dominant distortion mode in terms of its contributions from Pb- and I-displacements, since the band edges and hence the size of the band gap of these materials are derived from Pb $s$ and I $p$ (VBM) and Pb $p$ (CBM) orbitals, respectively.  We calculate the percentage of the total amplitude, $a_{total}$, corresponding to displacements of Pb and I, $a_{PbI} = a_{Pb} + a_{I}$, as $\sqrt{\sum a_{PbI}^2}/a_{total} \cdot a_{PbI}$. This percentage contribution is shown in color in Figures~\ref{fig:gaps}(a), (c), and (e), respectively.

Our analysis shows that the influence of the mode on the band gap depends on the Pb-I contribution of the mode. For modes where the amplitude of the dominant mode leads to large band gap changes, as exemplified by our first case in Figure~\ref{fig:gaps}(a), the contribution is almost constant at 100\%. For our second case in Figure~\ref{fig:gaps}(c), the dominant mode amplitude is $\sim$1\,\AA{} for a wide range of different structures with varying degrees of secondary and tertiary mode contributions. However, structures with dominant $X_3^+$ contribution primarily coming from Pb-I distortions have larger band gaps than those structures in which the distortion also contains significant Cs contributions. In the third case of Figure~\ref{fig:gaps}(e), the lowest band gap structure with $Imm2$ symmetry is derived entirely from a $\Gamma_5^-$ mode with 100\% PbI character (an off-center displacement of Pb), but the amplitude of the mode is very small. The $\Gamma_5^-$ mode of the other structures has a significantly lower PbI contribution, indicating that their large band gap originates from their secondary $X_3^-$ or $X_4^-$ modes which have amplitudes between 1.1 and 1.5\,\AA{}. 

\section{Magnitude of Rashba-/Dresselhaus Splitting in Polar Structures} \label{sec:Rashba}
\begin{figure}[ht]
\centering
\includegraphics[width=0.7\linewidth]{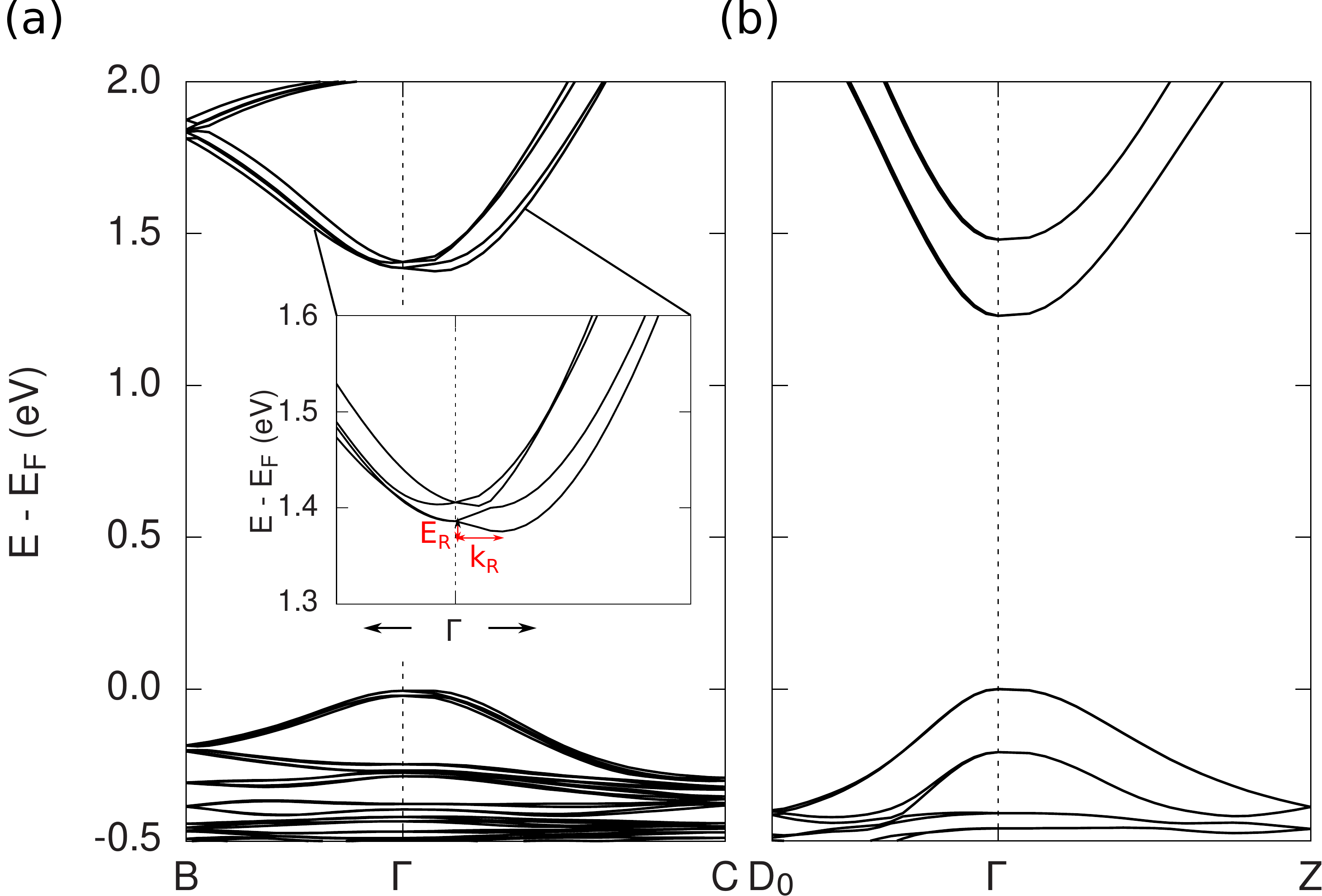}
\caption{\label{fig:bandstructure}Band structures calculated using DFT-PBE+SOC. (a) $n=1$ \cspbiRpSingle{} in polar $Pc$ symmetry.  The inset shows the conduction band minimum and our definition of the Rashba energy E$_R$ and momentum k$_R$. (b) $n=2$ \cspbiRpDouble{} in polar $Cmc2_1$ symmetry.}
\label{fig:Rashba}
\end{figure}

We now focus on the band structure of \cspbiRpSingle{} and \cspbiRpDouble{} perovskites featuring a polar space group. Polar, i.e., non-centrosymmetric materials, with strong SOC in their electronic structure can exhibit Rashba- \cite{Rashba1960}, Dresselhaus- \cite{Dresselhaus1955}, or mixed Rashba-/Dresselhaus-splitting. In all three cases, the defining characteristic is a splitting of the energy bands in $\mathbf k$-space, that we quantify by the parameter $\alpha = 2E_R/k_R$, with E$_R$ and k$_R$ as defined in Figure~\ref{fig:Rashba}(a). In the following, we will not distinguish between pure and mixed Rashba and/or Dresselhaus cases and refer to the effect as Rashba-/Dresselhaus splitting. An in-depth exploration of these spin-textures can be found in Ref.~\cite{Maurer2022}.

We start by calculating the macroscopic electronic polarization of all polar \cspbiRpSingle{} and \cspbiRpDouble{} structures using the Berry phase approach and with the Born effective charge approximation. These values are given in Table~\ref{tab:polarizationRashba} and are in very good agreement with each other. For \cspbiRpSingle{} only the structures with $Pc$ and $Aea2$ symmetry display a significant polarization ($>1\,\mu C/cm^2$) of 3.4 and 2.7\,$\mu C/cm^2$, respectively. For \cspbiRpDouble{}, the ground state structure with $Cmc2_1$ symmetry has a polarization of 1.9\,$\mu C/cm^2$. The two other phases with significant polarization are $Aem2$ with 1.8\,$\mu C/cm^2$ and $Pm$ with 1.6\,$\mu C/cm^2$, respectively. Note that these values are of the same order of magnitude as what was calculated for all-inorganics 3D ABX$_3$ perovskites \cite{Leppert2016}, but well below the polarization of typical oxide ferroelectrics.

\begin{table}[ht]
\centering
\caption{Macroscopic polarization ($\mu C/cm^2$) computed according to the modern theory of polarization (p$_{Berry}$) and the Born effective charge approximation (p$_{BEC}$). Rashba-/Dresselhaus $\alpha$ parameter calculated as described in the main text in eV\AA{}. Large $\alpha$ parameters highlighted in green.}
\label{tab:polarizationRashba}
\begin{tabular}{llllllll}
\br
\multicolumn{4}{c}{$n=1$}  & \multicolumn{4}{c}{$n=2$}\\%mr
space group & p$_{Berry}$ & p$_{BEC}$ & $\alpha$ & space group & p$_{Berry}$ & p$_{BEC}$ & $\alpha$ \\ \mr
Pmc2$_1$    & 0.03        & 0.03      & 0.0      & Cmc2$_1$    & 1.91        & 1.93      & 0.0      \\
Cc          & 0.02        & 0.02      & 0.0      & Cm          & 0.04        & 0.03      & 0.0      \\
P1          & 0.01        & 0.01      & 0.0      & Pmn2$_1$    & 0.08        & 0.09      & 0.0      \\
P2$_1$      & 0.00        & 0.01      & 0.0      & Pmm2        & 0.01        & 0.01      & 0.0      \\
C2          & 0.04        & 0.04      & 0.0      & P2          & 0.02        & 0.01      & 0.0      \\
\mycc Pc          & \mycc 3.44        & \mycc 3.20      & \mycc 0.9     & Aem2        & 1.75        & 1.78      & 0.0      \\
\mycc Pmn2$_1$    &  \mycc 0.35        & \mycc 0.41      & \mycc 0.9     & P2$_1$      & 0.04        & 0.07      & 0.0      \\
\mycc Aea2        & \mycc 2.66        & \mycc 2.66      & \mycc 0.7     & P1          & 0.85        & 0.86      & 0.0      \\
Imm2        & 0.80        & 0.80      & 0.0      & Pma2        & 0.01        & 0.01      & 0.0      \\
I4mm        & 0.12        & 0.12      & 0.0      & Pc          & 0.78        & 0.77      & 0.0      \\
Fmm2        &     0.38        &     0.38      &   0.0       & Ama2        & 0.14        & 0.12      & 0.0 \\
            &             &           &          & Amm2        & 0.11        & 0.09      & 0.0      \\
            &             &           &          & C2          & 0.25        & 0.24      & 0.0     \\
            &             &           &          & \mycc Pm          & \mycc 1.57        & \mycc 1.37      & \mycc 0.5     \\
            &             &           &          & Pmc2$_1$    & 0.85        & 0.84      & 0.0      \\
            &             &           &          & I4mm        & 0.01        & 0.12      & 0.0      \\
            &             &           &          & Imm2        & 0.73        & 0.73      & 0.0      \\
            &             &           &          & Fmm2        & 2.10        & 1.81   &  0.0 \\\br
\end{tabular}
\end{table}

For evaluating the magnitude of the Rashba-/Dresselhaus effect, we calculated the band structure of all polar structures listed in Table~\ref{tab:polarizationRashba} across the entire Brillouin zone including SOC self-consistently. The parameter $\alpha$ was calculated by fitting a fourth-degree polynomial to the conduction band. For $Aea2$ we additionally evaluated the $\alpha$ parameter using a regular three-dimensional grid around the $\Gamma$ point to ensure that we are measuring the splitting in the reciprocal space direction where it is largest. The value reported in Table~\ref{tab:polarizationRashba} refers to this calculation. Note that due to numerical accuracy, a Rashba-/Dresselhaus effect smaller than $\sim$0.1\,eV\AA{} will not be resolved in our calculations.

Our main finding is that even though a total of 27 structures have polar space groups in our study, a clear Rashba-/Dresselhaus effect at the CBM, i.e., around the $\Gamma$ point is only visible in four structures: $Pmn2_1$, $Aea2$, and $Pc$ for \cspbiRpSingle{} and $Pm$ for \cspbiRpDouble{}. For many of the analyzed structures this finding can simply be explained by the vanishing polarization of the structures. However, for some structures with significant polarization, for example the polar ground state of \cspbiRpDouble{} and the $Aem2$ structure, we also do not observe Rashba-/Dresselhaus splitting of the CBM or in the lowest-energy conduction band at other $\mathbf k$-points across the Brillouin zone. As shown below, this finding can be explained by the direction and Pb-I orbital contribution of the polar distortion mode.

In \cspbiRpSingle{}, all three structures with significant Rashba-/Dresselhaus splitting are dominated by a $\Sigma_3$ mode. In these structures, the initial octahedral tilting pattern results in enough symmetry breaking to allow for the appearance of the $\Gamma^-_5$ polar mode. Similarly, for \cspbiRpDouble{}, the $Pm$ structure with sizeable Rashba-/Dresselhaus effect is dominated by a $\Sigma_1$ mode. We note that the interoctahedral asymmetries studied by Jana \textit{et al.} may also result in the appearance of the $\Gamma^-_5$ polar mode. However, as opposed to the case of quasi-2D hybrid perovskites, where octahedral tilt angle asymmetries in the inorganic monolayer arise out of distortions at the organic-inorganic interface, in all-inorganic RP perovskites, the $\Gamma^-_5$ polar mode arises from a combination of coherent mode distortions, that lead to a global polar symmetry. Notably, in the $n=2$ structures $Cmc2_1$ and $Aem2$, Rashba-/Dresselhaus splitting is not observed (see Figure~\ref{fig:bandstructure}(b)) because the polar $\Gamma_5^-$ mode primarily comprises Cs ion displacements which do not contribute electronically to the band edges. Finally, our results are also in agreement with Ref.~\cite{Maurer2022}, since all the structures displaying relatively large splitting contain polar atomic displacements towards an out-of-bond direction of the octahedron. However, we note that in equilibrium, $n=1$ and $n=2$ $Fmm2$ structures display small polarizations and negligible splitting effects despite their out-of-bond distortion. 

\section{Energy Landscape and Polarization of Free-Standing 2D Halide Perovskite Layers}\label{sec:2D}
Pb-based all-inorganic RP perovskites have at this point and to the best of our knowledge not been synthesized. However, a multitude of hybrid quasi-2D perovskites exist \cite{Smith2018a}. Additionally, these materials can be prepared as single crystal monolayers by mechanical exfoliation \cite{Yaffe2015a}, prompting us to ask the question whether our findings for RP halide perovskites can be translated to mono- and bilayer halide perovskites. For this purpose we constructed mono- and bilayer slices from fully relaxed CsPbI$_3$ and separated periodic repeat units by 20\,\AA{} of vacuum to avoid spurious interlayer interactions. The undistorted structure of this material class has $P4/mmm$ symmetry. Inducing all possible tilt patterns to the reference structure as before results in 13 and 16 unique space groups for the mono- and bilayer case, respectively. Importantly, and in contrast to the RP perovskites, octahedral rotations cannot break inversion symmetry in the monolayer case. For the bilayer, we find four polar space groups. A list of space groups, their energies relative to the reference structure, and their DFT-PBE band gaps can be found in \ref{sec:appendix3}.

The ground state structures for the mono- and bilayer have $P\bar{1}$ and $Cmmm$ symmetry and energy gains of -105\,meV and -163\,meV with respect to the reference structure, respectively. Similar to before, we find a clear energetic grouping of structures based on their dominant distortion mode. For the monolayer, structures with a dominant $M_5^+$ distortion, corresponding to an $a^-a^0a^0$ tilt pattern, i.e., out-of-phase rotation along the $a$- and $b$-directions, have the lowest energy, similar to the RP halide perovskites. This is followed by two structures with dominant $M_3^+$ mode ($a^0a^0a^+$) and another cluster with dominant $X_3^+$ ($a^+a^0a^0$) mode. This is different from the RP perovskites for which the in-phase rotation around the $c$-direction was the energetically least favorable. For the bilayer, the energetic ordering in terms of dominant distortion modes is the same as for the $n=2$ RP perovskites. In particular, the polar $Pmc2_1$ structure which is dominated by a large $\Gamma_5^-$ mode is originating from an alternating shift of Cs ions which is responsible for a significant polarization of this structure. 

In line with our results of Section~\ref{sec:Rashba}, none of the polar bilayer structures exhibits Rashba-/Dresselhaus splitting, since in all cases structures with a significant polarization are dominated by either an in-bond $\Gamma_5^-$ mode or by a $\Gamma_5^-$ mode associated primarily with Cs displacements.
    
\section{Discussion and Summary}
\label{sec:finalsection}

Fully inorganic RP halide perovskites are challenging to synthesize, and in particular, \cspbiRpSingle{} and \cspbiRpDouble{} have not been experimentally realized yet. However, an $n=1$ RP perovskite Cs$_2$PbI$_2$Cl$_2$ with corner-sharing octahedra in which I occupies the out-of-plane halide sites was reported by Li \textit{et al.} in Ref.~\cite{li_cs2pbi2cl2_2018}. Furthermore, Yu \textit{et al.} reported the formation of a RP phase in nanosheets of CsPbBr$_3$, suggesting routes for stabilizing the RP phase through nanostructuring \cite{yu_ruddlesdenpopper_2017}. Since DFT calculations predict many all-inorganic RP halide perovskite compositions to be thermodynamically unstable \cite{li_cs2pbi2cl2_2018}, it is important to connect our findings to the much more common hybrid organic-inorganic RP halide perovskites. The energy landscape and structural distortions in these hybrid systems are determined by steric effects and interactions at the organic-inorganic interface, allowing for a wider range of distortions and other structural parameters such as the interlayer distance and stacking. Nonetheless, since band gaps and the Rashba-/Dresselhaus effect are determined by the inorganic sublattice, we expect our findings for the effect of octahedral tilts on the band structure and the emergence of Rashba-/Dresselhaus splitting to be of relevance for the hybrid systems as well.

We have here focused on RP halide perovskites with $n=1$ and $n=2$. The main difference between these systems is that the extra layer in the $n=2$ system allows for a coherent octahedral tilting in the perovskite plane. The latter leads to a larger set of unique octahedral tilting distortions and more possibilities to induce polar distortions due to the combination of symmetry-breaking nonpolar octahedral rotations. For $n>2$, we expect an increase in the cooperativity of the octahedral tilting in both in-plane and out-of-plane directions. The number and overall characteristics of the possible octahedral tilt patterns will remain similar due to the RP fault separating the perovskites layers. As the number of layers $n$ increases, we expect the interior layers, in the middle of the perovskite slab, to approach the rotation angle in the 3D perovskite $n=\infty$. The perovskite layers near the fault may experience strain and smaller rotation angles, closer to the $n=2$ values, due the restrictions of the RP geometry. The latter suggests that, as $n$ increases, properties such as polarization, will become increasingly dominated by the perovskite slab as opposed to the layers near the fault. We expect polar structures with in-plane polarization to approach bulk 3D polar structures, with equivalent band gaps and polarization.

Rashba-/Dresselhaus splitting is expected for any structure without inversion symmetry and large spin-orbit interaction. In our case, Rashba-/Dresselhaus splitting is observed for polar structures with significant polarization. However, our work emphasizes that besides these basic ingredients, Rashba-/Dresselhaus splitting requires polar atomic displacements to be associated with band edge orbitals.
While the computationally predicted Rashba-/Dresselhaus splitting is larger for 3D perovskite or large $n$ with dominant perovskite layer over the RP fault, polar structures are increasingly less stable for $n \rightarrow \infty$. Therefore, the observation of the Rashba-/Dresselhaus effect is increasingly less likely for large $n$. In contrast, whereas Rashba-/Dresselhaus splitting is smaller for the possible polar structures with $n=1$ and $n=2$, these polar structures are close to the ground state and therefore more likely to be observed experimentally. In this context, we emphasize that the $n=1$ and $n=\infty$ systems are fundamentally different from RP perovskites with finite $n>1$, and deserve separate analysis. For finite $n$, the RP fault inevitably leads to multiple octahedral tiltings, and Rashba-/Dresselhaus splitting analogous to $n=2$ as studied in our work and by Maurer et al for $n=2,3$ \cite{Maurer2022}. With increasing $n$, we expect the perovskite layers in the RP structure to dominate and converge to the values of 3D perovskites.

We note that octahedral tilting distortions are ubiquitous in halide perovskites (as well as in oxide perovskites and related materials). In 3D perovskites, these distortions are not polar and cannot induce ferroelectricity \cite{Young2016}. As is well-known for oxide and shown explicitly here for halide perovskites, layering in RP structures can lead to a sizable electric polarization due to the coupling of rotational modes \cite{Benedek2011}. RP halide perovskites therefore also have potential applications as ferroelectrics if a significant switchable electric polarization can be achieved \cite{zhang_ferroelectricity_2022}. But also in photovoltaic applications a macroscopic electric polarization can be beneficial due to its connection to the bulk photovoltaic effect \cite{yang_above-bandgap_2010}. Furthermore, since polar structures with strong spin-orbit coupling can feature a sizable Rashba-/Dresselhaus splitting, applications in spin-orbitronics can also be envisioned \cite{Kepenekian2017}. Finally, RP structures can be used to achieve quantum confinement effects in halide perovskites \cite{Reyes-Lillo2016}. 

In summary, we have systematically studied the complex energetic and electronic structure landscape of all-inorganic RP halide perovskites. Octahedral tilts allow for band gap changes comparable to those associated with halide substitution in these materials. We find a multitude of polar structures for both $n=1$ and $n=2$ RP halide perovskites; contrary to 3D ABX$_3$ perovskites polar space groups are accessible via octahedral tilts in RP perovskites. However, many of these polar structures have only very small polarization and only four display a significant Rashba-/Dresselhaus splitting in the CBM due to the direction and Pb-I contribution of their dominant polar distortion mode. This situation is even more pronounced in mono- and bilayers of these materials, since the lack of a "stacked" or layered structure as present in the RP perovskites reduces the amount of possibilities for generating polar structures via octahedral tilts.

Experimentally, targeted structural distortions could be induced through layering with organic molecules as has been demonstrated for a variety of RP halide perovskites \cite{Jana2020, Jana2021, chakraborty_rational_2023}. Another powerful tool for tailoring structural properties of perovskites is strain induced by epitaxial growth. For oxides, strain has been shown to dramatically change the energy landscape \cite{dieguez_first-principles_2005, Reyes-Lillo2019} and lead to functional properties \cite{piamonteze_interfacial_2015, hallsteinsen_concurrent_2016, xu_reducing_2018}. Recently, targeted epitaxial growth has been reported for halide perovskites \cite{Chen2020}, opening new routes to tune optoelectronic properties for photovoltaic applications and beyond.

\newpage
\appendix
\section{Data for Other Halides with n=1}\label{sec:appendix1}
This Appendix provides a Table with tilt patterns, energies, and band gaps of the $n=1$ bromide and chloride RP perovskites not reported in the main text, as well as the phonon band structure of \cspbiRpSingle{} computed as described in Section~\ref{sec:methods}.
\begin{figure}[hb]
\centering
\includegraphics[width=0.8\linewidth]{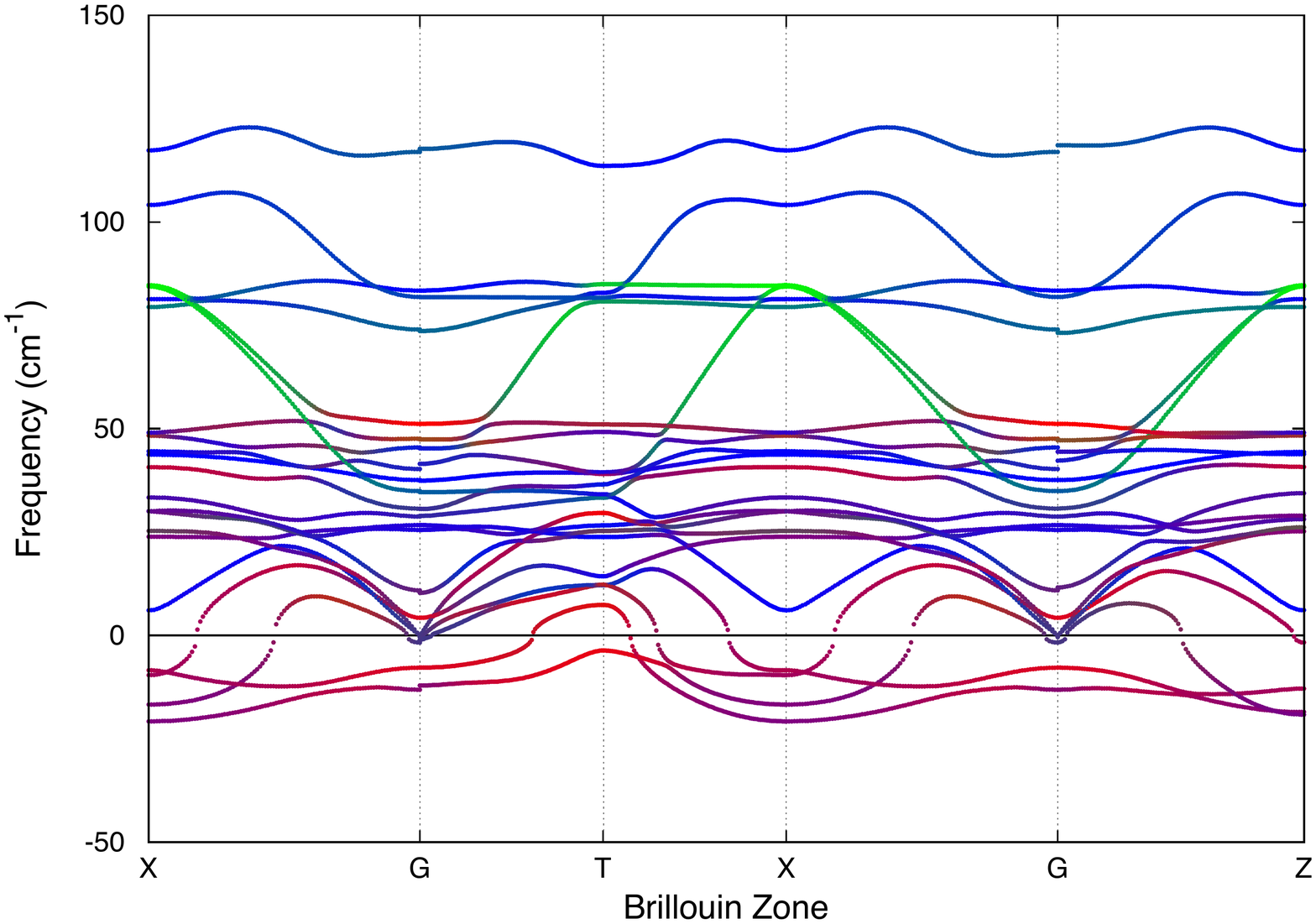}
\caption{\label{fig:phonon_bands_n=1} Phonon band structure of the $I4/mmm$ structure of \cspbiRpSingle{}. The colors of the bands indicate the contributions of Cs (red), Pb (green) and I (blue) displacements to each phonon mode.}
\end{figure}
\newpage

\begin{table}[ht]
  \centering
  \caption{Initial tilt pattern, energy gain $\Delta$E, size and position of the band gap for  \cspbbrRpSingle{} and \cspbclRpSingle{}. The tilt pattern and position of the band gap are the same for both  materials. All values were calculated using PBE, without SOC. Polar space groups are marked in green.}
  \resizebox*{0.8\textwidth}{!}{
    \begin{tabular}{rllllrrrr}
    \br
          &       &       &       &       & \multicolumn{2}{c}{\cspbbrRpSingle{}} & \multicolumn{2}{c}{\cspbclRpSingle{}} \\
    \multicolumn{1}{l}{space group} &       & \multicolumn{2}{c}{tilt pattern} & $\mathbf k$-point & \multicolumn{1}{l}{$\Delta$E (meV)} & \multicolumn{1}{l}{band gap (eV)} & \multicolumn{1}{l}{$\Delta$E (meV)} & \multicolumn{1}{l}{band gap (eV)} \\\mr
    \rowcolor[rgb]{ .776,  .878,  .706} 1     & P1    & a$^-$a$^+$a$^0$/a$^-$a$^+$a$^+$ & 111   & $\Gamma$ & -48.64 & 2.35 & -43.11 & 2.78 \\
    2     & P\minus{1}   & a$^0$b$^-$b$^+$/a$^0$b$^-$b$^+$ & 11-1  & Z     & -49.81 & 2.35 & -44.14 & 2.77 \\
    \rowcolor[rgb]{ .776,  .878,  .706} 4     & P2$_1$ & a$^+$a$^-$a$^+$/a$^+$a$^-$a$^+$ & 111   & $\Gamma$ & -49.25 & 2.35 & -43.84 & 2.77 \\
    \rowcolor[rgb]{ .776,  .878,  .706} 5     & C2    & a$^-$a$^+$a$^+$/a$^+$a$^-$a$^+$ & 111   & $\Gamma$ & -31.89 & 2.38 & -25.69 & 2.77 \\
    \rowcolor[rgb]{ .776,  .878,  .706} 6     & Pm    & a$^-$a$^+$a$^+$/a$^0$a$^+$a$^0$ & 111   & $\Gamma$ & -55.70 & 2.39 & -50.65 & 2.84 \\
    \rowcolor[rgb]{ .776,  .878,  .706} 7     & Pc    & a$^+$b$^+$b$^+$/a$^+$b$^+$b$^+$ & 11-1  & $\Gamma$ & -16.42 & 2.45 & -11.46 & 2.87 \\
    \rowcolor[rgb]{ .776,  .878,  .706} 9     & Cc    & a$^-$a$^+$a$^+$/a$^+$a$^-$a$^+$ & 1-11  & $\Gamma$ & -55.77 & 2.39 & -50.63 & 2.84 \\
    10    & P2/m  & a$^-$a$^0$a$^+$/a$^0$a$^+$a$^0$ & 111   & $\Gamma$ & -55.89 & 2.39 & -50.79 & 2.84 \\
    11    & P2$_1$/m & a$^-$a$^0$a$^0$/a$^0$a$^+$a$^+$ & 111   & $\Gamma$ & -55.96 & 2.39 & -50.85 & 2.84 \\
    12    & C2/m  & a$^-$a$^-$a$^0$/a$^+$a$^+$a$^0$ & 111   & $\Gamma$ & -47.94 & 2.32 & -42.97 & 2.73 \\
    13    & P2/c  & a$^-$b$^-$c$^0$/a$^-$b$^-$c$^0$ & 1-11  & $\Gamma$ & -49.59 & 2.35 & -43.23 & 2.79 \\
    14    & P2$_1$/c & a$^0$b$^-$b$^+$/a$^0$b$^-$b$^+$ & 111   & Y     & -47.03 & 2.47 & -41.98 & 2.85 \\
    15    & C2/c  & a$^0$a$^-$c$^+$/a$^0$a$^-$c$^+$ & 111   & $\Gamma$ & -27.61 & 2.45 & -21.61 & 2.87 \\
    18    & P2$_1$2$_1$2 & a$^-$a$^+$a$^0$/a$^+$a$^0$a$^0$ & 111   & $\Gamma$ & -55.38 & 2.39 & -50.11 & 2.84 \\
    \rowcolor[rgb]{ .776,  .878,  .706} 26    & Pmc2$_1$ & a$^-$a$^+$a$^0$/a$^0$a$^+$a$^0$ & 111   & $\Gamma$ & -54.51 & 2.39 & -50.65 & 2.84 \\
    \rowcolor[rgb]{ .776,  .878,  .706} 31    & Pmn2$_1$ & a$^0$a$^+$a$^0$/a$^0$a$^+$a$^+$ & 111   & $\Gamma$ & -13.06 & 2.41 & -9.49 & 2.83 \\
    \rowcolor[rgb]{ .776,  .878,  .706} 41    & Aea2  & a$^+$a$^+$a$^+$/a$^+$a$^+$a$^+$ & 11-1  & $\Gamma$ & -5.22 & 2.38 & -4.43 & 2.79 \\
    51    & Pmma  & a$^-$a$^0$a$^0$/a$^0$a$^+$a$^0$ & 111   & $\Gamma$ & -55.94 & 2.39 & -50.74 & 2.84 \\
    53    & Pmna  & a$^-$a$^-$a$^0$/a$^0$a$^0$a$^0$ & 111   & $\Gamma$ & -48.75 & 2.32 & -43.44 & 2.72 \\
    55    & Pbam  & a$^0$a$^0$a$^+$/a$^0$a$^0$c$^+$ & 111   & $\Gamma$ & 1.89  & 2.51 & -0.04 & 2.95 \\
    56    & Pccn  & a$^-$a$^-$c$^+$/a$^-$a$^-$c$^+$ & 111   & $\Gamma$ & -29.97 & 2.47 & -24.08 & 2.88 \\
    57    & Pbcm  & a$^-$a$^0$a$^0$/a$^+$a$^-$a$^0$ & 111   & $\Gamma$ & -31.84 & 2.36 & -25.86 & 2.80 \\
    59    & Pmmn  & a$^0$a$^0$a$^0$/a$^0$a$^+$a$^+$ & 111   & $\Gamma$ & 0.79  & 2.27 & -0.25 & 2.71 \\
    61    & Pbca  & a$^-$a$^-$c$^+$/a$^-$a$^-$c$^+$ & -1-1-1 & $\Gamma$ & -50.01 & 2.48 & -44.42 & 2.90 \\
    62    & Pnma  & a$^0$a$^+$c$^0$/a$^0$a$^+$c$^0$ & 111   & X     & -12.28 & 2.40 & -8.48 & 2.84 \\
    64    & Cmce  & a$^0$a$^0$c$^+$/a$^0$a$^0$c$^+$ & 111   & $\Gamma$ & 3.12  & 2.55 & 0.95  & 2.99 \\
    66    & Cccm  & a$^-$a$^-$c$^0$/a$^-$a$^-$c$^0$ & 111   & $\Gamma$ & -29.40 & 2.28 & -23.33 & 2.68 \\
    67    & Cmme  & a$^-$a$^0$a$^0$/a$^0$a$^0$a$^0$ & 111   & $\Gamma$ & -56.61 & 2.39 & -51.18 & 2.84 \\
    85    & P4/n  & a$^0$a$^0$a$^+$/a$^+$a$^+$a$^0$ & 111   & $\Gamma$ & 1.38  & 2.36 & -0.02 & 2.80 \\
    86    & P4$_2$/n & a$^-$a$^-$c$^0$/a$^-$a$^-$c$^0$ & 1-11  & $\Gamma$ & -56.61 & 2.39 & -51.08 & 2.84 \\
    94    & P4$_2$2$_1$2 & a$^-$a$^+$a$^0$/a$^+$a$^-$a$^0$ & 111   & $\Gamma$ & -31.28 & 2.38 & -26.02 & 2.80 \\
    114   & P\minus{4}2$_1$c & a$^-$a$^+$a$^0$/a$^+$a$^-$a$^0$ & 1-11  & $\Gamma$ & -55.81 & 2.39 & -50.65 & 2.84 \\
    127   & P4/mbm & a$^0$a$^0$a$^0$/a$^0$a$^0$a$^+$ & 111   & $\Gamma$ & 1.41  & 2.27 & 0.43  & 2.71 \\
    129   & P4/nmm & a$^0$a$^0$a$^0$/a$^+$a$^+$a$^0$ & 111   & $\Gamma$ & 1.21  & 2.27 & 0.01  & 2.71 \\
    134   & P4$_2$/nnm & a$^-$a$^0$a$^0$/a$^0$a$^-$a$^0$ & 111   & $\Gamma$ & -33.18 & 2.36 & -27.03 & 2.80 \\
    137   & P4$_2$/nmc & a$^0$a$^+$a$^0$/a$^+$a$^0$a$^0$ & 111   & $\Gamma$ & -0.71 & 2.34 & -1.20 & 2.78 \\
    138   & P4$_2$/ncm & a$^-$a$^0$a$^0$/a$^0$a$^-$a$^0$ & 1-11  & $\Gamma$ & -56.58 & 2.39 & -51.09 & 2.84 \\
    139   & I4/mmm & a$^0$a$^0$a$^0$/a$^0$a$^0$a$^0$ & 111   & $\Gamma$     & 0.00  & 2.27 & 0.00  & 2.71 \\
    \rowcolor[rgb]{ .776,  .878,  .706} 42    & Fmm2  & a$^0$a$^0$a$^0$/a$^0$a$^0$a$^0$ & 111   &  M  &  0.00    & 2.27 & -0.01 & 2.71   \\
    \rowcolor[rgb]{ .776,  .878,  .706} 44    & Imm2  & a$^0$a$^0$a$^0$/a$^0$a$^0$a$^0$ & 111      & M     & -0.28 & 2.31 & -0.11 & 2.76 \\
    \rowcolor[rgb]{ .776,  .878,  .706} 107   & I4mm  & a$^0$a$^0$a$^0$/a$^0$a$^0$a$^0$ & 111       & $\Gamma$     & -0.05 & 2.27 & 0.01  & 2.72 \\\br
   \end{tabular}
  \label{tab:all}%
  }
\end{table}%

\newpage
\section{Data for Other Halides with n=2}\label{sec:appendix2}
The following Table contains tilt patterns, energies, and band gaps for the bromide and chloride $n=2$ RP perovskites not reported in the main text.
\begin{table}[ht]
  \centering
  \caption{Initial tilt pattern, energy gain $\Delta$E, size and position of the band gap for  \cspbbrRpDouble{} and \cspbclRpDouble{}. The tilt pattern and position of the band gap are the same for both materials. All values were calculated using PBE, without SOC. Polar space groups are marked in green.}
  \resizebox*{0.7\textwidth}{!}{
    \begin{tabular}{rllllrrrr}
    \br
          &       &       &       &       & \multicolumn{2}{c}{\cspbbrRpDouble{}} & \multicolumn{2}{c}{\cspbclRpDouble{}} \\
    \multicolumn{1}{l}{space group} &       & \multicolumn{2}{c}{tilt pattern} & $\mathbf k$-point & \multicolumn{1}{l}{$\Delta$E (meV)} & \multicolumn{1}{l}{band gap (eV)} & \multicolumn{1}{l}{$\Delta$E (meV)} & \multicolumn{1}{l}{band gap (eV)} \\\mr
    \rowcolor[rgb]{ .776,  .878,  .706} 1     & P1    & a$^-$a$^-$a$^-$/a$^-$a$^-$a$^+$ & 1-11  & $\Gamma$ & -81.32 & 2.30 & -72.51 & 2.71 \\
    2     & P\minus{1}   & a$^0$b$^-$b$^-$/a$^0$b$^-$b$^-$ & 111   & Z     & -96.31 & 2.21 & -86.08 & 2.64 \\
    \rowcolor[rgb]{ .776,  .878,  .706} 3     & P2    & a$^-$a$^0$a$^-$/a$^0$a$^+$a$^+$ & 111   & $\Gamma$ & -94.65 & 2.24 & -87.04 & 2.67 \\
    \rowcolor[rgb]{ .776,  .878,  .706} 4     & P21   & a$^+$a$^-$a$^-$/a$^+$a$^-$a$^-$ & 11-1  & $\Gamma$ & -80.91 & 2.29 & -85.24 & 2.64 \\
    \rowcolor[rgb]{ .776,  .878,  .706} 5     & C2    & a$^-$a$^-$a$^-$/a$^+$a$^+$a$^-$ & 111   & $\Gamma$ & -82.42 & 2.27 & -76.71 & 2.67 \\
    \rowcolor[rgb]{ .776,  .878,  .706} 6     & Pm    & a$^+$b$^+$b$^+$/a$^+$b$^+$b$^+$ & 11-1  & $\Gamma$ & -95.50 & 2.21 & -85.26 & 2.64 \\
    \rowcolor[rgb]{ .776,  .878,  .706} 7     & Pc    & a$^-$a$^-$a$^-$/a$^-$a$^-$a$^+$ & 111   & $\Gamma$ & -70.85 & 2.29 & -62.36 & 2.73 \\
    \rowcolor[rgb]{ .776,  .878,  .706} 8     & Cm    & a$^0$b$^-$b$^+$/a$^0$b$^-$b$^+$ & 111   & $\Gamma$ & -96.74 & 2.21 & -86.04 & 2.65 \\
    10    & P2/m  & a$^-$b$^-$c$^0$/a$^-$b$^-$c$^0$ & 1-11  & $\Gamma$ & -96.75 & 2.21 & -86.22 & 2.65 \\
    11    & P21/m & a$^+$a$^-$c$^0$/a$^+$a$^-$c$^0$ & 111   & $\Gamma$ & -95.04 & 2.21 & -85.15 & 2.64 \\
    12    & C2/m  & a$^0$a$^-$c$^0$/a$^0$a$^-$c$^0$ & 111   & $\Gamma$ & -96.67 & 2.21 & -86.13 & 2.64 \\
    13    & P2/c  & a$^-$a$^-$a$^-$/a$^-$a$^-$a$^-$ & 111   & $\Gamma$ & -61.46 & 2.30 & -55.20 & 2.70 \\
    14    & P21/c & a$^0$b$^-$b$^-$/a$^0$b$^-$b$^-$ & 11-1  & $\Gamma$ & -96.58 & 2.21 & -86.03 & 2.64 \\
    15    & C2/c  & a$^-$a$^-$a$^-$/a$^-$a$^-$a$^-$ & -1-1-1 & $\Gamma$ & -80.26 & 2.30 & -74.40 & 2.69 \\
    16    & P222  & a$^0$a$^0$a$^-$/a$^0$a$^+$a$^+$ & 111   & $\Gamma$ & -24.84 & 2.20 & -25.46 & 2.64 \\
    17    & P222$_1$ & a$^0$a$^0$a$^+$/a$^0$a$^+$a$^-$ & 111   & $\Gamma$ & -27.23 & 2.24 & -26.61 & 2.68 \\
    20    & C2221 & a$^+$a$^+$a$^-$/a$^+$a$^+$a$^-$ & 111   & $\Gamma$ & -47.83 & 2.28 & -42.17 & 2.68 \\
    21    & C222  & a$^0$a$^0$a$^-$/a$^+$a$^+$a$^-$ & 111   & $\Gamma$ & -23.77 & 2.27 & -24.87 & 2.70 \\
    \rowcolor[rgb]{ .776,  .878,  .706} 25    & Pmm2  & a$^-$a$^0$a$^0$/a$^0$a$^+$a$^+$ & 111   & $\Gamma$ & -94.91 & 2.24 & -87.14 & 2.67 \\
    \rowcolor[rgb]{ .776,  .878,  .706} 26    & Pmc2$_1$ & a$^-$a$^-$a$^0$/a$^-$a$^-$a$^+$ & 111   & $\Gamma$ & -67.23 & 2.26 & -60.07 & 2.66 \\
    \rowcolor[rgb]{ .776,  .878,  .706} 28    & Pma2  & a$^-$a$^0$a$^+$/a$^0$a$^-$a$^+$ & 111   & $\Gamma$ & -74.56 & 2.24 & -65.70 & 2.67 \\
    \rowcolor[rgb]{ .776,  .878,  .706} 31    & Pmn2$_1$ & a$^-$a$^0$a$^+$/a$^0$a$^-$a$^+$ & 1-11  & $\Gamma$ & -96.31 & 2.24 & -87.85 & 2.68 \\
    \rowcolor[rgb]{ .776,  .878,  .706} 36    & Cmc2$_1$ & a$^-$a$^-$a$^+$/a$^-$a$^-$a$^+$ & -1-1-1 & $\Gamma$ & -102.22 & 2.35 & -90.41 & 2.78 \\
    \rowcolor[rgb]{ .776,  .878,  .706} 38    & Amm2  & a$^-$a$^-$a$^0$/a$^+$a$^+$a$^0$ & 111   & $\Gamma$ & -77.87 & 2.23 & -72.76 & 2.62 \\
    \rowcolor[rgb]{ .776,  .878,  .706} 39    & Aem2  & a$^-$a$^-$a$^+$/a$^-$a$^-$a$^+$ & 111   & $\Gamma$ & -82.91 & 2.34 & -71.13 & 2.76 \\
    \rowcolor[rgb]{ .776,  .878,  .706} 40    & Ama2  & a$^+$a$^+$a$^+$/a$^+$a$^+$a$^+$ & 11-1  & $\Gamma$ & -72.35 & 2.25 & -63.57 & 2.68 \\
    47    & Pmmm  & a$^-$a$^0$a$^0$/a$^0$a$^+$a$^0$ & 111   & $\Gamma$ & -95.48 & 2.24 & -87.32 & 2.68 \\
    49    & Pccm  & a$^0$a$^0$a$^-$/a$^0$a$^+$a$^0$ & 111   & $\Gamma$ & -25.06 & 2.19 & -25.91 & 2.64 \\
    50    & Pban  & a$^0$a$^0$a$^-$/a$^0$a$^0$c$^-$ & 111   & $\Gamma$ & -25.82 & 2.39 & -27.68 & 2.81 \\
    51    & Pmma  & a$^-$a$^-$a$^0$/a$^0$a$^0$a$^0$ & 111   & $\Gamma$ & -79.52 & 2.23 & -74.22 & 2.62 \\
    54    & Pcca  & a$^-$a$^-$a$^-$/a$^-$a$^-$a$^-$ & 11-1  & $\Gamma$ & -64.59 & 2.28 & -57.77 & 2.68 \\
    55    & Pbam  & a$^0$a$^0$a$^+$/a$^0$a$^0$c$^+$ & 111   & $\Gamma$ & -24.65 & 2.37 & -26.04 & 2.78 \\
    57    & Pbcm  & a$^-$a$^-$a$^+$/a$^-$a$^-$a$^+$ & 11-1  & $\Gamma$ & -78.79 & 2.33 & -67.55 & 2.76 \\
    59    & Pmmn  & a$^0$a$^+$c$^0$/a$^0$a$^+$c$^0$ & 111   & X     & -45.74 & 2.23 & -39.10 & 2.66 \\
    60    & Pbcn  & a$^-$a$^-$a$^-$/a$^-$a$^-$a$^-$ & -1-11 & $\Gamma$ & -83.82 & 2.30 & -77.37 & 2.70 \\
    62    & Pnma  & a$^-$a$^-$a$^+$/a$^-$a$^-$a$^+$ & -1-11 & $\Gamma$ & -97.29 & 2.34 & -86.24 & 2.77 \\
    63    & Cmcm  & a$^-$a$^-$c$^0$/a$^-$a$^-$c$^0$ & -1-11 & $\Gamma$ & -79.34 & 2.23 & -73.97 & 2.63 \\
    64    & Cmce  & a$^0$a$^0$c$^+$/a$^0$a$^0$c$^+$ & 111   & $\Gamma$ & -24.33 & 2.38 & -25.85 & 2.79 \\
    65    & Cmmm  & a$^-$a$^0$a$^0$/a$^0$a$^0$a$^0$ & 111   & $\Gamma$ & -96.34 & 2.24 & -87.38 & 2.68 \\
    66    & Cccm  & a$^-$a$^+$a$^0$/a$^+$a$^-$a$^0$ & 111   & $\Gamma$ & -71.02 & 2.26 & -62.54 & 2.68 \\
    67    & Cmme  & a$^-$a$^-$c$^0$/a$^-$a$^-$c$^0$ & 111   & $\Gamma$ & -60.24 & 2.22 & -53.62 & 2.60 \\
    68    & Ccce  & a$^0$a$^0$c$^-$/a$^0$a$^0$c$^-$ & 111   & $\Gamma$ & -25.56 & 2.40 & -27.54 & 2.82 \\
    81    & P\minus{4}   & a$^0$a$^0$a$^+$/a$^+$a$^+$a$^-$ & 111   & $\Gamma$ & -25.05 & 2.28 & -24.43 & 2.69 \\
    83    & P4/m  & a$^0$a$^0$a$^+$/a$^+$a$^+$a$^0$ & 111   & $\Gamma$ & -21.80 & 2.24 & -23.08 & 2.67 \\
    84    & P42/m & a$^-$a$^-$c$^0$/a$^-$a$^-$c$^0$ & 1-11  & $\Gamma$ & -96.16 & 2.24 & -87.38 & 2.68 \\
    89    & P422  & a$^0$a$^0$a$^-$/a$^+$a$^+$a$^0$ & 111   & $\Gamma$ & -23.40 & 2.24 & -23.57 & 2.67 \\
    115   & P\minus{4}m2 & a$^0$a$^0$a$^0$/a$^+$a$^+$a$^-$ & 111   & $\Gamma$ & -10.31 & 2.11 & -11.31 & 2.52 \\
    117   & P\minus{4}b2 & a$^0$a$^0$a$^-$/a$^0$a$^0$a$^+$ & 111   & $\Gamma$ & -25.12 & 2.38 & -26.76 & 2.80 \\
    123   & P4/mmm & a$^0$a$^0$a$^0$/a$^+$a$^+$a$^0$ & 111   & $\Gamma$ & -10.35 & 2.10 & -11.36 & 2.52 \\
    125   & P4/nbm & a$^0$a$^0$a$^-$/a$^0$a$^0$a$^0$ & 111   & $\Gamma$ & -3.94 & 2.10 & -6.82 & 2.52 \\
    127   & P4/mbm & a$^0$a$^0$a$^0$/a$^0$a$^0$a$^+$ & 111   & $\Gamma$ & -3.48 & 2.10 & -6.25 & 2.52 \\
    131   & P4$_2$/mmc & a$^0$a$^+$a$^0$/a$^+$a$^0$a$^0$ & 111   & $\Gamma$ & -29.81 & 2.20 & -28.47 & 2.63 \\
    132   & P4$_2$/mcm & a$^-$a$^0$a$^0$/a$^0$a$^-$a$^0$ & 111   & $\Gamma$ & -74.44 & 2.24 & -65.55 & 2.67 \\
    136   & P42/mnm & a$^-$a$^0$a$^0$/a$^0$a$^-$a$^0$ & 1-11  & $\Gamma$ & -96.28 & 2.24 & -87.73 & 2.68 \\
    139   & I4/mmm & a$^0$a$^0$c$^0$/a$^0$a$^0$a$^0$ & 111   & $\Gamma$     & 0.00  & 2.10 & 0.00  & 2.52 \\
    \rowcolor[rgb]{ .776,  .878,  .706} 42    & Fmm2  & a$^0$a$^0$c$^0$/a$^0$a$^0$a$^0$ & 111      &  M    & -0.09 & 2.10 & -0.02  & 2.52  \\
    \rowcolor[rgb]{ .776,  .878,  .706} 44    & Imm2  & a$^0$a$^0$c$^0$/a$^0$a$^0$a$^0$ & 111       & M     & -0.18 & 2.12 & 0.02  & 2.55 \\
    \rowcolor[rgb]{ .776,  .878,  .706} 107   & I4mm  & a$^0$a$^0$c$^0$/a$^0$a$^0$a$^0$ & 111       & $\Gamma$     & 0.11  & 2.10 & 0.07  & 2.52 \\\br
     \end{tabular}
  \label{tab:bromines}%
  }
\end{table}%

\newpage
\section{Data for Free-Standing Mono- and Bilayer Perovskites}\label{sec:appendix3}
The following Table contains tilt patterns, energies, and band gaps of the mono- and bilayer perovskites of Section~\ref{sec:2D}.
\begin{table}[ht]
\caption{Initial tilt pattern, energy gain $\Delta$E, size and position of the band gap for 2D mono- and bilayer structures. Polar groups are marked in green. }
\centering
%\begin{subtable}[t]{0.49\linewidth}
\resizebox*{0.5\linewidth}{!}{
    \begin{tabular}[t]{llrcc}
    \br
    \multicolumn{5}{c}{$n=1$ \cspbiRpSingle{}} \\ %\mr
    space group & tilt pattern & $\mathbf k$-point & \multicolumn{1}{c}{$\Delta$E} (meV) & \multicolumn{1}{l}{band gap} (eV) \\ \mr
    P\minus{1}   & a$^-$b$^-$a$^+$ & Z & -104.57 & 2.16 \\
    P2$_1$/m & a$^-$a$^+$a$^+$ & $\Gamma$ & -96.99 & 1.98 \\
    C2/m  & a$^-$a$^0$a$^+$ & $\Gamma$ & -99.90 & 1.94 \\
    P2/c  & a$^-$b$^-$a$^0$ & $\Gamma$ & -96.57 & 1.95 \\
    P2$_1$/c & a$^-$a$^-$a$^+$ & $\Gamma$ & -97.62 & 2.13 \\
    Pmma  & a$^0$a$^+$a$^0$ & X & -10.59 & 1.90 \\
    Pmna  & a$^-$a$^-$a$^0$ & Z & -91.32 & 1.94 \\
    Pbcm  & a$^-$a$^+$a$^0$ & $\Gamma$ & -97.12 & 1.98 \\
    Pmmn  & a$^0$a$^+$a$^+$ & $\Gamma$ & -33.76 & 2.23 \\
    Cmme  & a$^-$a$^0$a$^0$ & $\Gamma$ & -100.15 & 1.94 \\
    P4/mmm & a$^0$a$^0$a$^0$ & M & 0.00  & 1.84 \\
    P4/mbm & a$^0$a$^0$a$^+$ & $\Gamma$ & -34.53 & 2.24 \\
    P4/nmm & a$^+$a$^+$a$^0$ & $\Gamma$ & -6.15 & 1.97  \\\br
    \end{tabular}
    }
\end{table}
%\begin{subtable}[t]{0.49\linewidth}
\begin{table}[ht]
\centering
\resizebox*{0.5\linewidth}{!}{
     \begin{tabular}[t]{llrcc}
    \br
    \multicolumn{5}{c}{$n=2$ \cspbiRpDouble{}} \\ %\mr
    space group & tilt pattern & $\mathbf k$-point & \multicolumn{1}{c}{$\Delta$E} (meV) & \multicolumn{1}{l}{band gap} (eV) \\ \mr
    P\minus{1}   & a$^-$b$^-$a$^-$  & $\Gamma$ & -162.45 & 1.86 \\
    \polColorRow Pm    & a$^-$b$^-$a$^+$  & $\Gamma$ & -132.38 & 1.92 \\
    P2/m  & a$^-$b$^-$a$^0$  & $\Gamma$ & -162.37 & 1.86 \\
    P2$_1$/m & a$^-$a$^+$a$^-$  & $\Gamma$ & -155.06 & 2.05 \\
    C2/m  & a$^-$a$^0$a$^-$  & $\Gamma$ & -162.38 & 1.86 \\
    P2/c  & a$^-$a$^-$a$^-$  & B & -124.77 & 1.98 \\
    \polColorRow Pmm2  & a$^-$a$^+$a$^+$  & $\Gamma$ & -160.55 & 1.87 \\
    \polColorRow Pmc2$_1$ & a$^-$a$^-$a$^+$  & $\Gamma$ & -157.30 & 2.03 \\
    \polColorRow Amm2  & a$^-$a$^0$a$^+$  & $\Gamma$ & -162.24 & 1.86 \\
    Pmmm  & a$^0$a$^+$a$^0$  & U & -57.54 & 1.82 \\
    Pmma  & a$^-$a$^-$a$^0$  & $\Gamma$ & -117.77 & 1.92 \\
    Cmmm  & a$^-$a$^0$a$^0$  & $\Gamma$ & -162.81 & 1.86 \\
    P\minus{4}m2 & a$^+$a$^+$a$^-$  & $\Gamma$ & -43.66 & 1.96 \\
    P4/mmm & a$^0$a$^0$a$^0$  & M & 0.00  & 1.76 \\
    P4/nbm & a$^0$a$^0$a$^-$  & $\Gamma$ & -37.89 & 2.07 \\
    P4/mbm & a$^0$a$^0$a$^+$  & $\Gamma$ & -36.63 & 2.07 \\\br
    \end{tabular}%
    }
%\end{subtable}
\label{tab:2D_tiltPattern}%
\end{table}

\section*{Acknowledgments}
We acknowledge computational resources provided by the Bavarian Polymer Center at the University of Bayreuth, Germany, the Dutch national supercomputing center Snellius supported by the SURF Cooperative and PRACE for awarding access to the Marconi100 supercomputer at CINECA, Italy. N.F.-C. and S.E.R.-L. acknowledges ANID Fondecyt Regular grant number 1220986. Powered@NLHPC: This research was partially supported by the supercomputing infrastructure of the NLHPC (ECM-02).

\section*{References}
\providecommand{\newblock}{}


\begin{thebibliography}{10}
\expandafter\ifx\csname url\endcsname\relax
  \def\url#1{{\tt #1}}\fi
\expandafter\ifx\csname urlprefix\endcsname\relax\def\urlprefix{URL }\fi
\providecommand{\eprint}[2][]{\url{#2}}

\bibitem{Mitzi1995}
Mitzi D~B, Wang S, Feild C~A, Chess C~A and Guloy A~M 1995 {\em Science\/} {\bf
  267} 1473
  \urlprefix\url{http://www.ncbi.nlm.nih.gov/pubmed/17743545}

\bibitem{Mitzi1996a}
Mitzi D~B 1996 {\em Inorg. Chem. Chem.\/} {\bf 1669} 7614--7619

\bibitem{saparov_organicinorganic_2016}
Saparov B and Mitzi D~B 2016 {\em Chem. Rev.\/} {\bf 116} 4558--4596
  \urlprefix\url{https://doi.org/10.1021/acs.chemrev.5b00715}

\bibitem{Tsai2016}
Tsai H, Nie W, Blancon J~C, Stoumpos C~C, Asadpour R, Harutyunyan B, Neukirch
  A~J, Verduzco R, Crochet J~J, Tretiak S, Pedesseau L, Even J, Alam M~A, Gupta
  G, Lou J, Ajayan P~M, Bedzyk M~J, Kanatzidis M~G and Mohite A~D 2016 {\em
  Nature\/} {\bf 536} 312
  \urlprefix\url{http://www.nature.com/doifinder/10.1038/nature18306}

\bibitem{Stoumpos2016a}
Stoumpos C~C, Cao D~H, Clark D~J, Young J, Rondinelli J~M, Jang J~I, Hupp J~T
  and Kanatzidis M~G 2016 {\em Chem. Mater.\/} {\bf 28} 2852--2867

\bibitem{Smith2018a}
Smith M~D, Crace E~J, Jaffe A and Karunadasa H~I 2018 {\em Ann. Rev. Mater.
  Res.\/} {\bf 48} 111--136
  \urlprefix\url{https://www.annualreviews.org/doi/10.1146/annurev-matsci-070317-124406}

\bibitem{Blancon2018}
Blancon J~C, Stier A~V, Tsai H, Nie W, Stoumpos C~C, Traoré B, Pedesseau L,
  Kepenekian M, Katsutani F, Noe G~T, Kono J, Tretiak S, Crooker S~A, Katan C,
  Kanatzidis M~G, Crochet J~J, Even J and Mohite A~D 2018 {\em Nature Comm.\/}
  {\bf 9} 2254
  \urlprefix\url{http://dx.doi.org/10.1038/s41467-018-04659-x}

\bibitem{Even2014}
Even J, Pedesseau L and Katan C 2014 {\em ChemPhysChem\/} {\bf 15} 3733--3741

\bibitem{mcnulty_structural_2021}
McNulty J~A and Lightfoot P 2021 {\em IUCrJ\/} {\bf 8} 485--513
  \urlprefix\url{https://journals.iucr.org/m/issues/2021/04/00/yc5031/}

\bibitem{Gao_MolecularEngineering_2019}
Gao Y, Shi E, Deng S, Shiring S~B, Snaider J~M, Liang C, Yuan B, Song R, Janke
  S~M, Liebman-Peláez A, Yoo P, Zeller M, Boudouris B~W, Liao P, Zhu C, Blum
  V, Yu Y, Savoie B~M, Huang L and Dou L 2019 {\em Nat. Chem.\/} {\bf 11} 1151
  -- 1157

\bibitem{Connor2018}
Connor B~A, Leppert L, Smith M~D, Neaton J~B and Karunadasa H~I 2018 {\em J.
  Am. Chem. Soc.\/} {\bf 140} 5235--5240
  \urlprefix\url{http://pubs.acs.org/doi/10.1021/jacs.8b01543}

\bibitem{Connor2020a}
Connor B~A, Biega R~I, Leppert L and Karunadasa H~I 2020 {\em Chem. Sci.\/}
  {\bf 11} 7708--7715
  \urlprefix\url{http://dx.doi.org/10.1039/D0SC01580F}

\bibitem{Coffey_ControllingCrystalization_2022}
Coffey A~H, Yang S~J, Gómez M, Finkenauer B~P, Terlier T, Zhu C and Dou L 2022
  {\em Adv. Energy Mat.\/}  2201501
  \urlprefix\url{https://onlinelibrary.wiley.com/doi/abs/10.1002/aenm.202201501}

\bibitem{Yaffe2015a}
Yaffe O, Chernikov A, Norman Z~M, Zhong Y, Velauthapillai A, van~der Zande A,
  Owen J~S and Heinz T~F 2015 {\em Phys. Rev. B\/} {\bf 92} 045414 \urlprefix\url{http://link.aps.org/doi/10.1103/PhysRevB.92.045414}

\bibitem{aubrey_directed_2021}
Aubrey M~L, Saldivar~Valdes A, Filip M~R, Connor B~A, Lindquist K~P, Neaton J~B
  and Karunadasa H~I 2021 {\em Nature\/} {\bf 597} 355--359
  \urlprefix\url{https://www.nature.com/articles/s41586-021-03810-x}

\bibitem{Shi_TwodimensionalHalide_2020}
Shi E, Yuan B, Shiring S~B, Gao Y, Akriti, Guo Y, Su C, Lai M, Yang P, Kong J,
  Savoie B~M, Yu Y and Dou L 2020 {\em Nature\/} {\bf 580} 614 -- 620

\bibitem{Pedesseau2016}
Pedesseau L, Sapori D, Traore B, Robles R, Fang H~H, Loi M~A, Tsai H, Nie W,
  Blancon J~C, Neukirch A, Tretiak S, Mohite A~D, Katan C, Even J and
  Kepenekian M 2016 {\em ACS Nano\/} {\bf 10} 9776--9786

\bibitem{cortecchia_structure-controlled_2018}
Cortecchia D, Neutzner S, Yin J, Salim T, Srimath~Kandada A~R, Bruno A, Lam
  Y~M, Martí-Rujas J, Petrozza A and Soci C 2018 {\em APL Materials\/} {\bf 6}
  114207
  \urlprefix\url{https://aip.scitation.org/doi/10.1063/1.5045782}

\bibitem{Fridriksson2020a}
Fridriksson M~B, van~der Meer N, de~Haas J and Grozema F~C 2020 {\em The
  Journal of Physical Chemistry C\/} {\bf 124} 28201--28209

\bibitem{Dhanabalan2020a}
Dhanabalan B, Leng Y~c, Bi G, Lin M~l, Tan P~h, Infante I, Manna L, Arciniegas
  M~P and Krahne R 2020 {\em ACS Nano\/} {\bf 14} 4689

\bibitem{Jana2020}
Jana M~K, Song R, Liu H, Khanal D~R, Janke S~M, Zhao R, Liu C, Valy~Vardeny Z,
  Blum V and Mitzi D~B 2020 {\em Nature Comm.\/} {\bf 11} 4699 ISSN 20411723
  publisher: Springer US
  \urlprefix\url{http://dx.doi.org/10.1038/s41467-020-18485-7}

\bibitem{Menahem2021}
Menahem M, Dai Z, Aharon S, Sharma R, Asher M, Diskin-Posner Y, Korobko R,
  Rappe A~M and Yaffe O 2021 {\em ACS Nano\/} {\bf 15} 10153--10162 ISSN
  1936086X

\bibitem{Li_2DRulebook_2021}
Li X, Hoffman J~M and Kanatzidis M~G 2021 {\em Chem. Rev.\/} {\bf 121}
  2230--2291
  \urlprefix\url{https://doi.org/10.1021/acs.chemrev.0c01006}

\bibitem{Filip2022}
Filip M~R, Qiu D~Y, Del~Ben M and Neaton J~B 2022 {\em Nano Lett.\/}
  \urlprefix\url{https://pubs.acs.org/doi/10.1021/acs.nanolett.2c01306}

\bibitem{mulder_turning_2013}
Mulder A~T, Benedek N~A, Rondinelli J~M and Fennie C~J 2013 {\em Adv.
  Funct. Mater.\/} {\bf 23} 4810--4820
  \urlprefix\url{https://onlinelibrary.wiley.com/doi/abs/10.1002/adfm.201300210}

\bibitem{Zhao2021}
Zhao X~G, Wang Z, Malyi O~I and Zunger A 2021 {\em Materials Today\/} {\bf 49}
  107--122
  \urlprefix\url{https://doi.org/10.1016/j.mattod.2021.05.021}

\bibitem{Benedek2013}
Benedek N~a and Fennie C~J 2013 {\em J. Phys. Chem. C\/} {\bf 117} 13339

\bibitem{Benedek2011}
Benedek N~A and Fennie C~J 2011 {\em Phys. Rev. Lett.\/} {\bf 106} 107204

\bibitem{Birol2011}
Birol T, Benedek N~A and Fennie C~J 2011 {\em Phys. Rev. Lett.\/} {\bf
  107} 257602

\bibitem{balachandran_crystal-chemistry_2014}
Balachandran P~V, Puggioni D and Rondinelli J~M 2014 {\em Inorg.
  Chem.\/} {\bf 53} 336--348

\bibitem{Even2015a}
Even J, Pedesseau L, Katan C, Kepenekian M, Lauret J~S, Sapori D and Deleporte
  E 2015 {\em J. Phys. Chem. C\/} 10161-10177
  \urlprefix\url{http://dx.doi.org/10.1021/acs.jpcc.5b00695}

\bibitem{Kepenekian2017}
Kepenekian M and Even J 2017 {\em J. Phys. Chem. Lett.\/} {\bf 8} 3362--3370
  \urlprefix\url{http://pubs.acs.org/doi/abs/10.1021/acs.jpclett.7b01015}

\bibitem{Rashba1960}
Rashba E 1960 {\em Sov. Phys. Solid State\/} {\bf 2} 1109--1122

\bibitem{Dresselhaus1955}
Dresselhaus G 1955 {\em Phys. Rev. B\/} {\bf 100} 580--586

\bibitem{Manchon2015}
Manchon A, Koo H~C, Nitta J, Frolov S~M and Duine R~a 2015 {\em Nature Mat.\/}
  {\bf 14} 871--882
  \urlprefix\url{http://www.nature.com/doifinder/10.1038/nmat4360}

\bibitem{Yang2016e}
Yang Y, Yang M, Zhu K, Johnson J~C, Berry J~J, van~de Lagemaat J and Beard M~C
  2016 {\em Nature Comm.\/} {\bf 7} 12613
  \urlprefix\url{http://www.nature.com/doifinder/10.1038/ncomms12613}

\bibitem{bourelle_optical_2022}
Bourelle S~A, Camargo F~V~A, Ghosh S, Neumann T, van~de Goor T~W~J, Shivanna R,
  Winkler T, Cerullo G and Deschler F 2022 {\em Nature Communications\/} {\bf
  13} 3320
  \urlprefix\url{https://www.nature.com/articles/s41467-022-30953-w}

\bibitem{Leppert2016}
Leppert L, Reyes-Lillo S~E and Neaton J~B 2016 {\em J. Phys. Chem. Lett.\/}
  {\bf 7} 3683--3689
  \urlprefix\url{https://pubs.acs.org/doi/full/10.1021/acs.jpclett.6b01794}

\bibitem{Zheng2015}
Zheng F, Tan L~Z, Liu S and Rappe A~M 2015 {\em Nano Lett.\/} {\bf 15}
  7794--7800
   \urlprefix\url{https://pubs.acs.org/doi/10.1021/acs.nanolett.5b01854}

\bibitem{Etienne2016a}
Etienne T, Mosconi E and De~Angelis F 2016 {\em J. Phys. Chem. Lett.\/} {\bf 7}
  1638--1645
  \urlprefix\url{http://dx.doi.org/10.1021/acs.jpclett.6b00564}

\bibitem{Bernardi2018}
Frohna K, Deshpande T, Harter J, Peng W, Barker B~A, Neaton J~B, Louie S~G,
  Bakr O~M, Hsieh D and Bernardi M 2018 {\em Nature Comm.\/} {\bf 9} 1829
  \urlprefix\url{http://www.nature.com/articles/s41467-018-04212-w}

\bibitem{Davies2018}
Davies C~L, Filip M~R, Patel J~B, Crothers T~W, Verdi C, Wright A~D, Milot R~L,
  Giustino F, Johnston M~B and Herz L~M 2018 {\em Nature Comm.\/} {\bf 9} 293
  \urlprefix\url{http://dx.doi.org/10.1038/s41467-017-02670-2}

\bibitem{Jana2021}
Jana M~K, Song R, Xie Y, Zhao R, Sercel P~C, Blum V and Mitzi D~B 2021 {\em
  Nature Comm.\/} {\bf 12} 4982 \urlprefix\url{http://dx.doi.org/10.1038/s41467-021-25149-7}

\bibitem{Maurer2022}
Maurer B, Vorwerk C and Draxl C 2022 {\em Phys. Rev. B\/} {\bf 105} 155149
 \urlprefix\url{https://journals.aps.org/prb/abstract/10.1103/PhysRevB.105.155149}
 
\bibitem{Kresse1996a}
Kresse G and Furthm\"uller J 1996 {\em Phys. Rev. B\/} {\bf 54}(16)
  11169--11186
  \urlprefix\url{https://link.aps.org/doi/10.1103/PhysRevB.54.11169}

\bibitem{Kresse1996b}
Kresse G and Furthmüller J 1996 {\em Comput. Mater. Sci.\/} {\bf
  6} 15--50
  \urlprefix\url{https://www.sciencedirect.com/science/article/pii/0927025696000080}

\bibitem{Kresse1999}
Kresse G and Joubert D 1999 {\em Phys. Rev. B\/} {\bf 59} 1758--1775 \urlprefix\url{http://link.aps.org/doi/10.1103/PhysRevB.59.1758}

\bibitem{Perdew1996}
Perdew J~P, Burke K and Ernzerhof M 1996 {\em Phys. Rev. Lett.\/} {\bf 77}
  3865--3868
  \urlprefix\url{http://www.ncbi.nlm.nih.gov/pubmed/10062328}

\bibitem{Campbell2006}
Campbell B~J, Stokes H~T, Tanner D~E and Hatch D~M 2006 {\em J. Appl. Cryst.\/}
  {\bf 39} 607--614 \urlprefix\url{https://doi.org/10.1107/S0021889806014075}

\bibitem{Togo2015}
Togo A and Tanaka I 2015 {\em Scripta Materialia\/} {\bf 108} 1--5
  \urlprefix\url{https://doi.org/10.1016/j.scriptamat.2015.07.021}

\bibitem{Giannozzi2009}
Giannozzi P, Baroni S, Bonini N, Calandra M, Car R, Cavazzoni C, Ceresoli D,
  Chiarotti G~L, Cococcioni M, Dabo I, {Dal Corso} A, {De Gironcoli} S, Fabris
  S, Fratesi G, Gebauer R, Gerstmann U, Gougoussis C, Kokalj A, Lazzeri M,
  Martin-Samos L, Marzari N, Mauri F, Mazzarello R, Paolini S, Pasquarello A,
  Paulatto L, Sbraccia C, Scandolo S, Sclauzero G, Seitsonen A~P, Smogunov A,
  Umari P and Wentzcovitch R~M 2009 {\em J. Phys.: Cond.
  Matt.\/} {\bf 21} 395502
  \urlprefix\url{https://iopscience.iop.org/article/10.1088/0953-8984/21/39/395502}

\bibitem{Giannozzi2017}
Giannozzi P, Andreussi O, Brumme T, Bunau O, {Buongiorno Nardelli} M, Calandra
  M, Car R, Cavazzoni C, Ceresoli D, Cococcioni M, Colonna N, Carnimeo I, {Dal
  Corso} A, {De Gironcoli} S, Delugas P, Distasio R~A, Ferretti A, Floris A,
  Fratesi G, Fugallo G, Gebauer R, Gerstmann U, Giustino F, Gorni T, Jia J,
  Kawamura M, Ko H~Y, Kokalj A, K{\"{u}}c{\"{u}}kbenli E, Lazzeri M, Marsili M,
  Marzari N, Mauri F, Nguyen N~L, Nguyen H~V, Otero-De-La-Roza A, Paulatto L,
  Ponc{\'{e}} S, Rocca D, Sabatini R, Santra B, Schlipf M, Seitsonen A~P,
  Smogunov A, Timrov I, Thonhauser T, Umari P, Vast N, Wu X and Baroni S 2017
  {\em J. Phys.: Cond. Matt.\/} {\bf 29} 465901
  \urlprefix\url{https://iopscience.iop.org/article/10.1088/1361-648X/aa8f79}

\bibitem{VanSetten2018}
van Setten M~J, Giantomassi M, Bousquet E, Verstraete M~J, Hamann D~R, Gonze X
  and Rignanese G~M 2018 {\em Comp. Phys. Comm.\/} {\bf 226}
  39--54

\bibitem{Deslippe2012}
Deslippe J, Samsonidze G, Strubbe D~A, Jain M, Cohen M~L and Louie S~G 2012
  {\em Comp. Phys. Comm.\/} {\bf 183} 1269--1289
  \urlprefix\url{http://dx.doi.org/10.1016/j.cpc.2011.12.006}

\bibitem{li_suppressing_2020}
Li S and Birol T 2020 {\em npj Comp. Mater.\/} {\bf 6} 168
  \urlprefix\url{https://www.nature.com/articles/s41524-020-00436-x}

\bibitem{zhang_strain-induced_2017}
Zhang Y, Sahoo M~P~K, Shimada T, Kitamura T and Wang J 2017 {\em Phys.
  Rev. B\/} {\bf 96} 144110
  \urlprefix\url{https://link.aps.org/doi/10.1103/PhysRevB.96.144110}

\bibitem{wang_first-principles_2021}
Wang C, Zhang M, Wang R, Zhang C, Meng X, Xi Y, Li S and Yan H 2021 {\em
  J. Phys. Chem. C\/} {\bf 125} 13971--13983
  \urlprefix\url{https://doi.org/10.1021/acs.jpcc.1c00720}

\bibitem{wang_ruddlesdenpopper_2016}
Wang H, Gou G and Li J 2016 {\em Nano Energy\/} {\bf 22} 507--513
  \urlprefix\url{https://www.sciencedirect.com/science/article/pii/S2211285516000872}

\bibitem{travis_application_2016}
Travis W, Glover E~N~K, Bronstein H, Scanlon D~O and Palgrave R~G 2016 {\em
  Chemical Science\/} {\bf 7} 4548--4556
  \urlprefix\url{https://pubs.rsc.org/en/content/articlelanding/2016/sc/c5sc04845a}

\bibitem{Filip2014}
Filip M~R, Eperon G~E, Snaith H~J and Giustino F 2014 {\em Nat. Comm.\/} {\bf
  5} 5757
  \urlprefix\url{http://www.nature.com/doifinder/10.1038/ncomms6757}

\bibitem{Wiktor2017a}
Wiktor J, Rothlisberger U and Pasquarello A 2017 {\em J. Phys. Chem. Lett.\/}
  {\bf 8} 5507--5512

\bibitem{Umari2013}
Umari P, Mosconi E and de~Angelis F 2014 {\em Sci. Rep.\/} {\bf
  4} 4467
  \urlprefix\url{https://www.nature.com/articles/srep04467}

\bibitem{Filip2014d}
Filip M~R and Giustino F 2014 {\em Phys. Rev B\/} {\bf 90} 245145 \urlprefix\url{http://link.aps.org/doi/10.1103/PhysRevB.90.245145}

\bibitem{Filip2016}
Filip M~R, Hillman S, Haghighirad A~A, Snaith H~J and Giustino F 2016 {\em J.
  Phys. Chem. Lett.\/} {\bf 7} 2579--2585
  \urlprefix\url{http://pubs.acs.org/doi/abs/10.1021/acs.jpclett.6b01041}

\bibitem{Leppert2019l}
Leppert L, Rangel T and Neaton J~B 2019 {\em Phys. Rev. Materials\/} {\bf 3}
  103803
  \urlprefix\url{https://journals.aps.org/prmaterials/abstract/10.1103/PhysRevMaterials.3.103803}

\bibitem{li_cs2pbi2cl2_2018}
Li J, Yu Q, He Y, Stoumpos C~C, Niu G, Trimarchi G~G, Guo H, Dong G, Wang D,
  Wang L and Kanatzidis M~G 2018 {\em J. Am. Chem.
  Soc.\/} {\bf 140} 11085--11090 \urlprefix\url{https://doi.org/10.1021/jacs.8b06046}

\bibitem{yu_ruddlesdenpopper_2017}
Yu Y, Zhang D and Yang P 2017 {\em Nano Lett.\/} {\bf 17} 5489--5494
  \urlprefix\url{https://doi.org/10.1021/acs.nanolett.7b02146}

\bibitem{Young2016}
Young J, Lalkiya P and Rondinelli J~M 2016 {\em J. Mater. Chem.
  C\/} {\bf 4} 4016--4027

\bibitem{zhang_ferroelectricity_2022}
Zhang Y, Parsonnet E, Fernandez A, Griffin S~M, Huyan H, Lin C~K, Lei T, Jin J,
  Barnard E~S, Raja A, Behera P, Pan X, Ramesh R and Yang P 2022 {\em Sci.
  Adv.\/} {\bf 8} eabj5881
  \urlprefix\url{https://www.science.org/doi/10.1126/sciadv.abj5881}

\bibitem{yang_above-bandgap_2010}
Yang S~Y, Seidel J, Byrnes S~J, Shafer P, Yang C~H, Rossell M~D, Yu P, Chu Y~H,
  Scott J~F, Ager J~W, Martin L~W and Ramesh R 2010 {\em Nature
  Nanotech.\/} {\bf 5} 143--147
  \urlprefix\url{https://www.nature.com/articles/nnano.2009.451}

\bibitem{Reyes-Lillo2016}
Reyes-Lillo S~E, Rangel T, Bruneval F and Neaton J~B 2016 {\em Phys. Rev.
  B\/} {\bf 94} 041107(R)
  \urlprefix\url{https://journals.aps.org/prb/abstract/10.1103/PhysRevB.94.041107}

\bibitem{chakraborty_rational_2023}
Chakraborty R, Rajput P~K, Anilkumar G~M, Maqbool S, Das R, Rahman A, Mandal P
  and Nag A 2023 {\em J. Am. Chem. Soc.\/} {\bf 145} 1378--1388
  \urlprefix\url{https://pubs.acs.org/doi/full/10.1021/jacs.2c12034}

\bibitem{dieguez_first-principles_2005}
Diéguez O, Rabe K~M and Vanderbilt D 2005 {\em Phys. Rev. B\/} {\bf 72}
  144101
  \urlprefix\url{https://link.aps.org/doi/10.1103/PhysRevB.72.144101}

\bibitem{Reyes-Lillo2019}
Reyes-Lillo S~E, Rabe K~M and Neaton J~B 2019 {\em Phys. Rev. Materials\/} {\bf
  3} 030601(R)
  \urlprefix\url{https://journals.aps.org/prmaterials/abstract/10.1103/PhysRevMaterials.3.030601}

\bibitem{piamonteze_interfacial_2015}
Piamonteze C, Gibert M, Heidler J, Dreiser J, Rusponi S, Brune H, Triscone J~M,
  Nolting F and Staub U 2015 {\em Phys. Rev. B\/} {\bf 92} 014426
  \urlprefix\url{https://link.aps.org/doi/10.1103/PhysRevB.92.014426}

\bibitem{hallsteinsen_concurrent_2016}
Hallsteinsen I, Moreau M, Grutter A, Nord M, Vullum P~E, Gilbert D~A, Bolstad
  T, Grepstad J~K, Holmestad R, Selbach S~M, N’Diaye A~T, Kirby B~J, Arenholz
  E and Tybell T 2016 {\em Phys. Rev. B\/} {\bf 94} 201115
  \urlprefix\url{https://link.aps.org/doi/10.1103/PhysRevB.94.201115}

\bibitem{xu_reducing_2018}
Xu R, Gao R, Reyes-Lillo S~E, Saremi S, Dong Y, Lu H, Chen Z, Lu X, Qi Y, Hsu
  S~L, Damodaran A~R, Zhou H, Neaton J~B and Martin L~W 2018 {\em ACS Nano\/}
  {\bf 12} 4736--4743
  \urlprefix\url{https://doi.org/10.1021/acsnano.8b01399}

\bibitem{Chen2020}
Chen Y, Lei Y, Li Y, Yu Y, Cai J, Chiu M~H, Rao R, Gu Y, Wang C, Choi W, Hu H,
  Wang C, Li Y, Song J, Zhang J, Qi B, Lin M, Zhang Z, Islam A~E, Maruyama B,
  Dayeh S, Li L~J, Yang K, Lo Y~H and Xu S 2020 {\em Nature\/} {\bf 577}
  209--215

\end{thebibliography}
\end{document}